\documentclass[nohyper]{JHEP3}

\usepackage{url}
\usepackage{epsfig}
\usepackage{latexsym}
\usepackage{amsmath}
\def\mathswitch#1{\relax\ifmmode#1\else$#1$\fi}

\newcommand{\sq}{\tilde{q}}

\usepackage{cancel}
\usepackage{dcolumn}
\newcolumntype{d}[0]{D{.}{.}{-1}}

\newcommand{\Slashed}[1]{\ensuremath{{#1}{\!}{\!}{\!}{\!}{\!}{\:}/}}

\newcommand{\eg}[0]{\textit{e.g.}}
\newcommand{\cf}[0]{\textit{cf.}}
\newcommand{\ie}[0]{\textit{i.e.}}

\newcommand{\neuino}[1]{\ensuremath{{\tilde{{\chi}}}_{#1}^{0}}}

\newcommand{\xone}[0]{\ensuremath{{\tilde{{\chi}}}_{1}^{0}}}

\newcommand{\mxone}[0]{\ensuremath{{\tilde{{\chi}}}_{1}^{0}}}
\newcommand{\go}[0]{\ensuremath{{\tilde{g}}}}
\newcommand{\se}[0]{\ensuremath{{\tilde{{\ell}}}}}

\newcommand{\Herwig}[0]{\texttt{Herwig++}}
\newcommand{\Cpp}[0]{\texttt{C++}}
\newcommand{\Fittino}[0]{\texttt{Fittino}}
\newcommand{\SPheno}[0]{\texttt{SPheno}}

\newcommand{\lsim}
{\;\raisebox{-.3em}{$\stackrel{\displaystyle <}{\sim}$}\;}
\newcommand{\gsim}
{\;\raisebox{-.3em}{$\stackrel{\displaystyle >}{\sim}$}\;}

\def\text{\textstyle}

\def\bc{\begin{center}}
\def\ec{\end{center}}
\def\bi{\begin{itemize}}
\def\ei{\end{itemize}}

\title{SUSY parameter determination at the LHC using cross sections
  and kinematic edges}

\author{Herbi K.~Dreiner\\
 Bethe Centre for Theoretical Physics \& Physikalisches Institut,
  Universit\"at Bonn, D-53115 Bonn, Germany and\\
  SCIPP, University of California, Santa Cruz, 95064 CA, USA}
\author{Michael Kr\"amer, Jonas M.~Lindert, Ben O'Leary\\
  Institut f\"ur Theoretische Teilchenphysik und Kosmologie,
  RWTH Aachen University,\\
  D-52056 Aachen, Germany}

\abstract{We study the determination of supersymmetric parameters at
  the LHC from a global fit including cross sections and edges of
  kinematic distributions. For illustration, we focus on a minimal
  supergravity scenario and discuss how well it can be constrained at
  the LHC operating at 7 and 14~TeV collision energy, respectively.
  We find that the inclusion of cross sections greatly improves the
  accuracy of the SUSY parameter determination, and allows to reliably
  extract model parameters even in the initial phase of LHC data
  taking with 7~TeV collision energy and 1~fb$^{-1}$ integrated
  luminosity. Moreover, cross section information may be essential to
  study more general scenarios, such as those with non-universal
  gaugino masses, and distinguish them from minimal, universal,
  models.}
  
\keywords{Supersymmetry Phenomenology, Hadronic Colliders}

\preprint{TTK-10-24}

\begin{document}

\section{Introduction}
\label{sec:intro}
Experiments at the Large Hadron Collider (LHC) will, for the first
time, directly explore supersymmetry (SUSY)~\cite{Golfand:1971iw,
  Wess:1974tw} at the Terascale.  Provided that evidence for
supersymmetry has been established, a major challenge will be to
determine the Lagrangian parameters of the theory, such as the SUSY
particle masses, their spins and couplings. These TeV-scale parameters
provide essential information on the scheme of supersymmetry breaking
and need to be determined with the highest possible accuracy.

A generic LHC signature for supersymmetric models with R-parity
conservation~\cite{Nilles:1983ge, Haber:1984rc} is that of cascade
decays of heavy squarks and gluinos which terminate in the lightest
supersymmetric particle (LSP).  In many SUSY models the LSP is weakly
interacting and provides a viable dark matter
candidate~\cite{Ellis:1983ew, Jungman:1995df}. However, a weakly
interacting LSP escapes detection and thus results in missing energy
signatures. It is a considerable challenge to reconstruct the
sparticle momenta in such cascade decays with missing energy at the
LHC and to determine sparticle masses and quantum numbers.  The
standard technique for analyzing SUSY cascade decays is to consider
invariant mass distributions of the final-state leptons and jets, see
\eg~\cite{Hinchliffe:1996iu, Allanach:2000kt, Gjelsten:2004ki,
  Weiglein:2004hn}. The kinematic endpoints of these distributions are
fixed by the masses of the sparticles in the decay chain and yield
model-independent information on part of the SUSY mass spectrum.  The
endpoints can be used as input for global SUSY pa\-ra\-me\-ter fits of
LHC data~\cite{Lester:2005je, Bechtle:2005vt, Lafaye:2007vs,
  Roszkowski:2009ye, Bechtle:2009ty} which determine the high-scale
model parameters and thereby test mechanisms of SUSY breaking.
(See~\cite{Allanach:2008zn} for a discussion of various fitting
methods and tools).  However, the endpoints of distributions are only
sensitive to the kinematics and can not constrain important SUSY
parameters such as $\tan\beta$ very well.  Moreover, as the endpoints
are determined by the difference of masses, there are ambiguities,
\ie\ points in the SUSY parameter space that have different spectra
but similar endpoints. There are further ambiguities from cascade
decays with identical final states, but different intermediate
particles. It is thus of vital importance to consider additional
information for the determination of SUSY parameters at the LHC.

In this paper we address the impact of event rates, \ie\ cross
sections and branching ratios, on SUSY pa\-ra\-meter fits of LHC data.
The cross sections are highly sensitive to the masses of the squarks
and gluinos that trigger the cascade decays. Branching ratios, on the
other hand, depend sensitively on the masses and mixings of SUSY
particles further down the decay chain. The inclusion of event rates
is technically challenging as the multi-dimensional parameter scan in
SUSY fits requires computationally extremely fast and reliable
theoretical estimates for cross sections, branching ratios and the
effect of experimental cuts.  The most straightforward and most
flexible approach of estimating event rates from a Monte Carlo
simulation for every point in the parameter fit is prohibitively
slow~\cite{Lester:2005je}, so that new strategies need to be
developed. Moreover, Monte Carlo calculations are generally based on
leading-order perturbation theory, resulting in a substantial
theoretical uncertainty in the prediction of event rates. Finally, the
statistical fluctuations of Monte Carlo estimates can cause
oscillations during the $\chi^2$-minimization in gradient based global
fits and lead to unstable results~\cite{Bechtle:2009ty}.

We propose a new approach to include event rates into SUSY parameter
fits. Our method is based on a simple parametrization of cross
sections and acceptances\footnote{We use the terms ``acceptance'' or
  ``cut acceptance'' to denote the fraction of events that passes the
  experimental selection cuts.  Our acceptance estimates are based on
  parton level calculations and do not yet include detector effects.},
and does not involve Monte Carlo simulations during the
$\chi^2$-minimization. It is thus fast and reproducible, and it
incorporates state-of-the-art higher-order cross section predictions
with small theoretical uncertainties.  Branching ratios, on the other
hand, can be evaluated very quickly with existing computational tools
(see \eg~\cite{Baer:2009tk}). The information on event rates, \ie\
cross sections$\,\times\,$branching ratios$\,\times\,$cut acceptances,
has been implemented as an additional observable into {\tt
  Fittino}~\cite{Bechtle:2004pc} and will become part of the next
official release of the {\tt Fittino} program
package~\cite{fittino_new}.

We expect event rate information to be particularly valuable for
general SUSY scenarios beyond minimal supergravity, for example those
which involve three-body decay modes, so that mass reconstruction via
the standard kinematic endpoints is difficult, see \eg\
\cite{Lester:2006cf, Baer:2007ya}. Moreover, parameter determinations
of theories with a general, non-universal, gaugino mass pattern may
play a crucial role to distinguish models of supersymmetry
breaking~\cite{Choi:2007ka}. As we shall see, cross sections and
branching ratios do not only add information to stabilize a fit with
the larger set of parameters in non-universal models, they also allow
to determine specific gaugino masses such as $M_3$ much more reliably.
Finally, production cross sections are essential to distinguish
supersymmetric theories from alternative new physics models, such as
universal extra dimensions~\cite{Appelquist:2000nn}, which predict
similar cascade decay signatures at the LHC, but involve new particles
with spin quantum numbers different from the SUSY particle
spectrum~\cite{Cheng:2002ab}. A universal extra dimension model could
not be distinguished from a supersymmetric theo\-ry based on the
kinematic information from endpoints of distributions, but would have
dramatically different cross sections because of the different
particle spins~\cite{Smillie:2005ar, Kane:2008kw, Hubisz:2008gg}.

To exemplify the impact of event rates on SUSY parameter fits we have
considered the standard SPS1a minimal supergravity (mSUGRA)
scenario~\cite{Allanach:2002nj}, which has been studied in great
detail in previous analyses. SPS1a provides a rich signature at the
LHC and can thus be constrained well by standard measurements of
kinematic endpoints. It is thus not a scenario where cross sections
and branching ratios are expected to have the largest impact.  On the
other hand, SPS1a allows for a quantitative analysis of the effects of
event rates on the parameter fit with realistic error estimates, as
discussed in detail in section~\ref{sec:numres}. Due to the relatively
light spectrum it is also one of the SUSY benchmark points that can be
studied in the initial phase of the LHC with 7~TeV collision energy
and low integrated luminosity.  Specifically, we investigate the
standard signatures of SPS1a-type models including two or more jets,
missing transverse momentum and two leptons of the same flavour but
opposite sign. Details of the cuts that specify the signatures will be
given in section~\ref{sec:cross_sections}. We use {\tt Fittino} to
determine the parameters of the SPS1a mSUGRA scenario from the
measurement of kinematic edges and event rates at the LHC, considering
7~TeV collision energy with 1~fb$^{-1}$ integrated luminosity and
14~TeV with both 1~fb$^{-1}$ and 10~fb$^{-1}$. As demonstrated in
section~\ref{sec:numres} the inclusion of rates in general stabilizes
the fit and significantly improves the error on the mSUGRA parameters,
in particular the universal gaugino mass $M_{1/2}$ and the ratio of
vacuum expectation values $\tan\beta$. Moreover, we find that the
inclusion of rates is crucial for the determination of mSUGRA
parameters in the initial phase of LHC data taking with 7~TeV
collision energy and 1~fb$^{-1}$ integrated luminosity. We also
consider a fit where we determine the gaugino mass parameters $M_1,
M_2, M_3$ individually at the GUT scale, instead of $M_{1/2}$, and
show that rate information is important to improve the accuracy of the
parameter determination for more general, non-universal, models.

The paper is structured as follows. Section~\ref{sec:cross_sections}
presents details of our method to estimate and parametrize the cross
sections and the effects of experimental cuts.  The SUSY parameter
fits are discussed in section~\ref{sec:numres}. We conclude in
section~\ref{sec:conclusion}.

\section{Cross sections and cut acceptances}
\label{sec:cross_sections}

The prediction of rates involves the calculation of the production
cross section for squarks and gluinos, which dominate the inclusive
SUSY signal at hadron colliders, the branching ratios of the
supersymmetric particles in the decay chains, and an estimate of the
effect of a certain set of experimental cuts. We have considered two
powerful and widely studied SUSY signatures~\cite{Hinchliffe:1996iu,
  Allanach:2000kt, Gjelsten:2004ki, Weiglein:2004hn}:
\begin{itemize}
\item[--] the inclusive signal of two or more hard jets with each
  $p_{T,{\rm jet}} > 50$ GeV, $|\eta| < 2.5$ as well as missing
  transverse energy of ${\Slashed{E}}_{T} > 100$ GeV;
\item[--] the exclusive signal of two opposite-sign same-flavour
  leptons ($e$ or $\mu$), each satisfying $p_{T,\ell} > 10$~GeV and
  $|\eta| < 2.5$, combined with the above signal of two or more hard
  jets and missing transverse energy. We remove the background from
  the leptonic decays of tau leptons, charginos and $W^{{\pm}}$ bosons
  in the standard way by subtracting events with opposite-sign
  different-flavour lepton pairs, see~\cite{Hinchliffe:1996iu,
    Allanach:2000kt, Gjelsten:2004ki, Weiglein:2004hn}.
\end{itemize}
Assuming a typical mSUGRA-like mass spectrum with $m_{\tilde{g}}>
m_{\tilde{q}}$, the hard jets result from two-body decays of gluinos
and squarks, $\tilde{g}\to q \tilde{q}$ and $\tilde{q}\to q
\tilde{\chi}$, respectively, where $\tilde{\chi}$ denotes a chargino
or neutralino, \ie\ a model-dependent linear combination of the
charged and neutral gauginos and higgsinos. The leptons are produced
in chargino/neutralino decays further down the cascade chain, \eg\
$\tilde{\chi}^0_2 \to \ell {\tilde\ell}_R \to \ell \ell
{\tilde\chi}^0_1$, as illustrated in Fig.~\ref{fig:cascade}.
\FIGURE{
  \includegraphics[width=0.5\textwidth,
  angle=0]{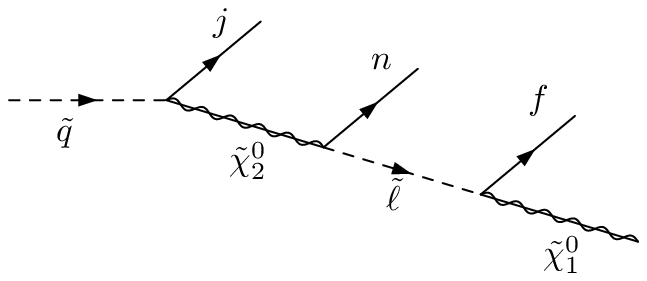}
  \caption{\label{fig:cascade} The cascade decay of a squark leading
    to the leptonic signal.  $j$ denotes the quark which should lead
    to a hard jet, $n$ denotes the near lepton, $f$ denotes the far
    lepton.}}
In the SPS1a scenario which we consider for illustration in this
paper, the neutralinos $\tilde{\chi}^0_{3,4}$ are mainly higgsino, so
that the decays $\tilde{q}\to q \tilde{\chi}^0_{3,4}$ are strongly
suppressed.  Decays into on-shell $Z$ or Higgs, ${\neuino{2}} \to Z
{\neuino{1}}$ or ${\neuino{2}} \to h {\neuino{1}}$, need a sufficient
$\tilde{\chi}^0_2-\tilde{\chi}^0_1$ mass splitting and are thus not
accessible for many mSUGRA scenarios, including SPS1a.  Such decays
into $Z$ or Higgs will be considered in the future, together with
other decay modes with potential significance for more general SUSY
models, but are not relevant for the results shown in this paper.

The calculation of the event rates proceeds in various stages. First,
for each point in the SUSY parameter space, the mass spectrum and
branching ratios are calculated by a spectrum generator. The masses
and branching ratios are highly model dependent, but they can be
evaluated quickly and can thus be calculated during the
$\chi^2$-minimization for every point in the parameter space.  Our
results are based on \SPheno~\cite{Porod:2003um}, which is the default
generator used by {\tt Fittino}.  Given the mass hierarchy, we first
check which cascade decays can contribute to the signal. The
corresponding event rates are proportional to the production cross
sections for squarks and gluinos.  These cross sections essentially
only depend on the squark and gluino masses and on no other SUSY
parameters or model assumptions.  They can thus be calculated once and
for all, including all available higher-order QCD corrections, stored
in a ($m_{\tilde{q}}, m_{\tilde{g}}$)-grid and read out quickly during
the fit, see section~\ref{sec:cross-sections}.  We calculate the cut
acceptances using a combination of numerical results stored in look-up
tables and analytical calculations of jet energies and lepton energies
in the squark rest frame. The calculation of the cut acceptances is
less straightforward, and details will be presented in
section~\ref{sec:cut_acceptances}.  For each contributing cascade, the
cross-section for the production of the relevant colored sparticles is
multiplied by the relevant branching ratios and the cut acceptance,
and the total event rate is passed on to {\tt Fittino} as an
observable entering the SUSY parameter fit.

In the following, we will describe the calculation of cross sections
and cut acceptances in some detail.

\subsection{Cross sections}
\label{sec:cross-sections}

In supersymmetric models with R-parity conservation, squarks and
gluinos are produced in pairs or associated pairs at hadron colliders,
$ pp \to \tilde{q}\tilde{q}, \tilde{q}\tilde{q}^*, \tilde{q}\tilde{g},
\tilde{g}\tilde{g} + X$.  Here we suppress the SU(2) quantum numbers
of the squarks $\tilde{q} =(\tilde{q}_{L}, \tilde{q}_{R})$ and do not
explicitly state the charge-conjugated processes. The squark and
gluino production cross sections are known including next-to-leading
order (NLO) SUSY-QCD corrections~\cite{Beenakker:1994an,
  Beenakker:1995fp, Beenakker:1996ch, Beenakker:1997ut}, the summation
of soft gluon emission~\cite{Kulesza:2008jb, Langenfeld:2009eg,
  Kulesza:2009kq, Beneke:2009rj, Beenakker:2009ha}, as well as
electroweak contributions and corrections~\cite{Bozzi:2005sy,
  Alan:2007rp, Bornhauser:2007bf, Hollik:2007wf, Hollik:2008yi,
  Hollik:2008vm, Mirabella:2009ap}.  The strength of the SUSY-QCD
interactions and thus the production rate is set by the gauge and the
Yukawa couplings of the $qqg$ and $q\tilde{q}\tilde{g}$ interactions,
respectively. The two couplings are required by supersymmetry to be
equal, so that the LO and NLO QCD squark and gluino parton cross
sections can be predicted unambiguously in terms of the QCD coupling
$g_{\rm s}$ and the squark and gluino masses, without any further SUSY
model dependence. The NLO-SUSY QCD corrections have been implemented
in the public computer code {\tt Prospino}~\cite{Beenakker:1996ed} and
form the basis of our cross section prediction. Note that the strong
Yukawa coupling between top quarks, top squarks and Higgs fields gives
rise to potentially large mixing effects and mass splitting in the top
squark sector. The mixing angle enters the top-squark cross section at
NLO, however, the dependence is numerically very
weak~\cite{Beenakker:1997ut}. Soft gluon resummation leads to a
reduction of the scale dependence of the cross section prediction and
an enhancement of the NLO QCD cross section for heavy squarks and
gluinos with masses $\tilde{m}\gsim 1$~TeV~\cite{Beenakker:2009ha}.
These effects will be included in future studies, but are not
essential for the results shown in this paper. Electroweak
contributions and corrections introduce a dependence on further
supersymmetric parameters. Their impact on the inclusive cross
sections summed over all squark species is generically small. However,
interference effects between the exchange of electroweak gauge bosons
and QCD contributions can become large for the production of two SU(2)
doublet squarks $pp \to \tilde{q}_L\tilde{q}'_L$~\cite{Alan:2007rp,
  Bornhauser:2007bf}.  For SPS1a-type scenarios, the electroweak
effects are moderate, but they may be larger for more general models,
in particular those without gaugino mass unification. Given the
residual QCD uncertainties of the production cross section discussed
below, we can neglect electroweak contributions for the study of SPS1a
presented in this paper.  However, such electroweak effects will be
addressed in future extensions of this work.

The calculation of the squark and gluino production cross sections
involves the calculation of the corresponding parton cross sections,
including higher-order SUSY-QCD corrections, and the convolution with
parton distribution functions. Numerically, this is a time-consuming
task, even at leading order, and would make the inclusion of cross
sections in SUSY parameter fits prohibitively slow.  Hence, we have
calculated the cross sections as a function of $m_{\tilde{q}}$ and
$m_{\tilde{g}}$ and stored the results in a grid. As the cross
sections are smooth functions of the sparticle masses, the values
between grid points can be interpolated reliably. When considering the
leptonic signature, we have to distinguish the SU(2) quantum numbers
of the squark, as decays from $\tilde{q}_R$ and $\tilde{q}_L$ are in
general very different. We have thus calculated the LO cross sections
for the production of squarks with definite SU(2) quantum number,
$\tilde{q}_L$ and $\tilde{q}_R$, averaging over the masses of
$\tilde{q} \in (\tilde{u}, \tilde{d}, \tilde{c}, \tilde{s})$ for each
SU(2) quantum number. Top and, to a lesser extend, bottom squarks mix
to form mass eigenstates and are thus treated separately, as mentioned
above and explained in detail in Ref.~\cite{Beenakker:1997ut}. The NLO
SUSY-QCD corrections are taken into account through $K$-factors,
$K\equiv \sigma_{\rm NLO}/\sigma_{\rm LO}$, as provided by {\tt
  Prospino}.  Note that the calculations~\cite{Beenakker:1994an,
  Beenakker:1995fp, Beenakker:1996ch} implemented in {\tt Prospino}
sum over squark SU(2) quantum numbers and do not provide separate
$K$-factors for the production of $\tilde{q}_L$ and $\tilde{q}_R$. We
thus assume that the $K$-factors do not depend significantly on the
squark SU(2) quantum number. Furthermore, we average the $K$-factors
for the $\tilde{q}\tilde{q}$ and $\tilde{q}\tilde{q}^*$ channels.  For
our numerical results, we have set the renormalization and
factorization scales to the average mass of the produced sparticles
and adopted the 2008 MSTW parton distribution
functions~\cite{Martin:2009iq}.

To illustrate the sensitivity of the SUSY cross section to the squark
and gluino masses, we show in Fig.~\ref{fig:NLO_cross-sections_14_TeV}
the NLO cross sections for $pp \to \tilde{q}\tilde{q},
\tilde{q}\tilde{q}^*, \tilde{q}\tilde{g}, \tilde{g}\tilde{g} + X$ at
the LHC with 14~TeV collision energy. In the Figure we sum over
$\tilde{q} \in (\tilde{u}, \tilde{d}, \tilde{c}, \tilde{s},
\tilde{b})$ and $L/R$ quantum numbers, and average over the squark
masses.

\vspace*{5mm}
\hspace*{125mm}$\sigma$~[pb]
\vspace*{-22mm}

\FIGURE{
  \includegraphics[width=0.7\textwidth,
  angle=-90]{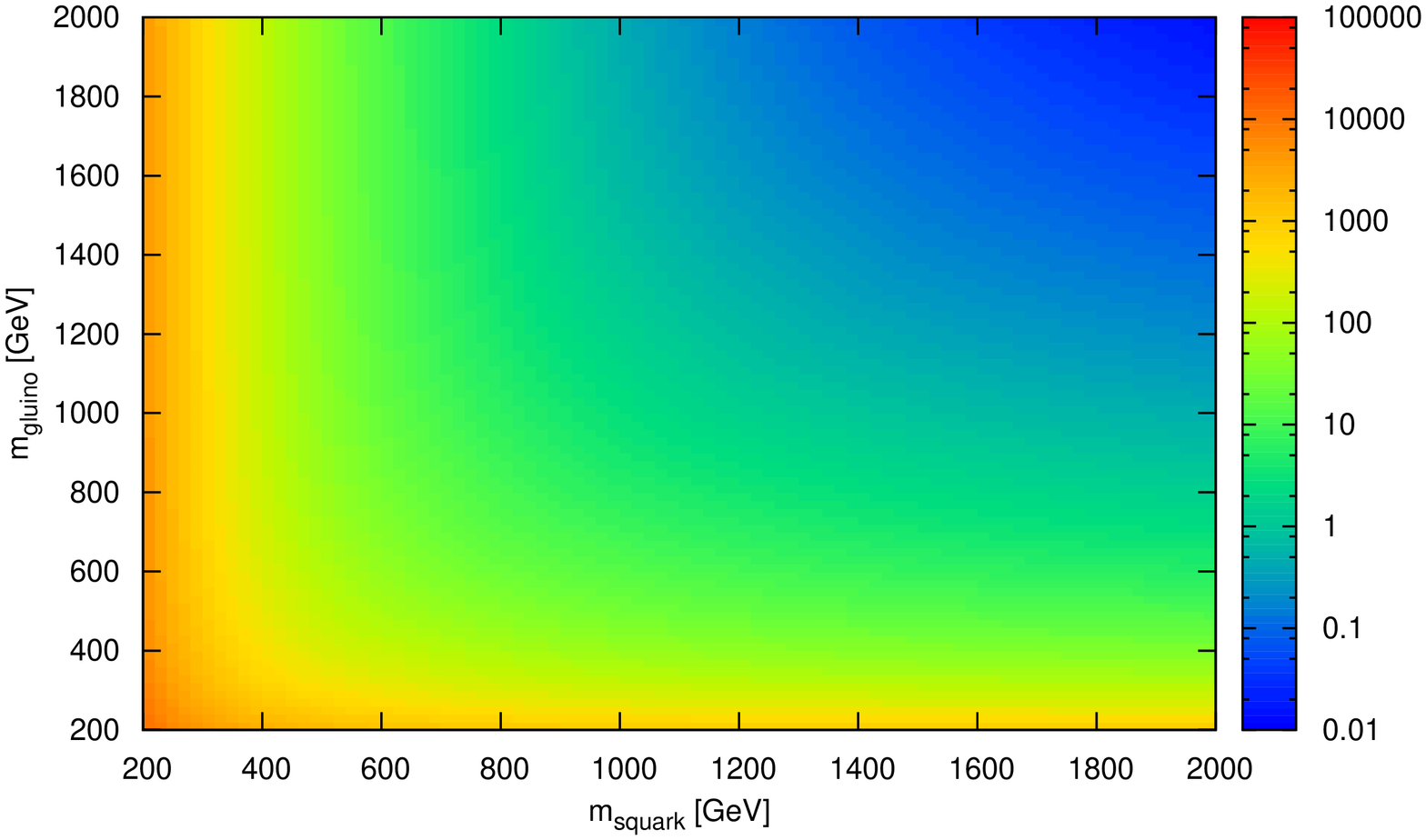}
  \caption{\label{fig:NLO_cross-sections_14_TeV} NLO QCD cross-section
    for inclusive squark and gluino production at the LHC (14~TeV) in
    pb, as a function of the gluino and average squark masses.}}
Varying ($m_{\tilde{q}}$, $m_{\tilde{g}}$) in the range between
200~GeV and 2~TeV, the cross section changes by seven orders of
magnitude. It is evident that the sensitivity of the cross section to
the sparticle masses should play an important role in SUSY parameter
fits.

Let us finally comment on the theoretical error of the cross section
prediction. The renormalization and factorization scale uncertainty of
the NLO QCD cross section is $\lsim 10\%$ for squark and gluino masses
below approximately 1~TeV~\cite{Beenakker:1996ch}. This uncertainty
could be reduced further by taking into account soft gluon
resummation~\cite{Beenakker:2009ha}. In addition, there is the
uncertainty due to the parton distribution functions, which, however,
is estimated to be below 5\% for sparticle masses less than about
1~TeV, see {\em e.g.}~\cite{Martin:2009iq, Nadolsky:2008zw,
  Ball:2010de}.  We thus assume an overall theoretical uncertainty of
15\% on our NLO cross section prediction. We shall discuss in
section~\ref{subsubse:Fittino_inputs} how this uncertainty enters the
SUSY parameter fit.

\subsection{Cut acceptances}
\label{sec:cut_acceptances}

The possibility of parameterizing the cut acceptances in a generic way
hinges upon factorizing the production of the colored sparticles and
the subsequent cascade decay. This factorization relies, of course, on
the narrow width approximation, which is appropriate as the width of
the SUSY particles is in general small, $\Gamma \ll
\tilde{m}$.\footnote{See Ref.~\cite{Berdine:2007uv} for new physics
  scenarios where finite width effects are important; these scenarios
  are, however, not relevant for the type of models studied in this
  paper.} We can thus break the problem of estimating cuts for the
full production$\,\otimes\,$decay process down to calculating the
decay distributions for jets and leptons analytically in the squark
rest frame and to numerically estimating the effect of boosting the
particles from the squark rest frame to the laboratory frame.  The
distributions of the squark decay products depend on the SUSY scenario
and on many SUSY parameters, but they can be calculated analytically,
and thus evaluated quickly during the SUSY parameter scan.  The impact
of the boost from the squark rest frame to the laboratory frame
depends on the dynamics of the production process and is in general
difficult to obtain analytically.  However, this part can be treated
numerically and stored in a look-up table, as it only depends on the
production dynamics and thus the squark and gluino masses, and not on
any specific details of the SUSY scenario. Our method will be outlined
briefly in sections~\ref{sec:jetcuts} and \ref{sec:leptoncuts}; more
details can be found in~\cite{JL}. Currently, the parametrization of
cut acceptances is implemented for signals that consist of sequences
of two-body decays with the usual mSUGRA mass hierarchy
$m_{\tilde{g}}> m_{\tilde{q}} > m_{\tilde{\chi}_2^0} >
m_{\tilde{\ell}} > m_{\tilde{\chi}_1^0}$. More general scenarios will
be considered in future work.

Even though the inclusive cross-sections for the production of colored
sparticles have large NLO QCD corrections in general, the NLO effects
on the differential distributions and thus on the typical boosts of
the colored sparticles are small~\cite{Beenakker:1996ch}.  Hence, for
the estimate of the acceptance we rely on leading order calculations.

\subsubsection{Jet and missing energy cuts}
\label{sec:jetcuts}

Unfortunately the missing energy cut is very difficult to calculate
analytically for a generic decay cascade, and we resorted to a
numerical grid of acceptances. The grid has been obtained by a simple
parton-level Monte Carlo simulation, decaying all particles by phase
space and ignoring spin correlations. This is a legitimate
approximation, as we average over charges in the final state.  We find
that the effects of intermediate decays from the squark to the
lightest neutralino tend to average out, and the missing energy cut
acceptance is approximated well as a function of the hard process
(giving the typical squark boosts) and just the mass difference
between the squark and the lightest neutralino. The acceptance grid
can thus be parametrized by three masses only: $m_{{\go}}$,
$m_{{\sq}}$ and $m_{\mxone}$.

Since all the jets from the decays of the gluinos and squarks were
simulated in the process of calculating the missing energy cut
acceptances, the jet cut acceptances were also taken from these
simulations, though in principle the jet cuts could have been dealt
with in the same manner as the lepton cuts described below. It is
intended to implement this in the future, to allow for the jet cuts to
be specified by the user, rather than hard-coded.

\subsubsection{Lepton cuts}
\label{sec:leptoncuts}

As argued above, we can estimate the impact of lepton cuts on the
production$\,\otimes\,$decay process by calculating the decay
distributions to leptons analytically in the squark rest frame and by
numerically estimating the effect of the squark boost to the
laboratory frame. We have derived analytical expressions for the
distributions of near and far leptons in the squark rest frame, which
can be evaluated very quickly during the global fit for every point in
parameter space. To estimate the lepton acceptance in the laboratory
frame, we numerically calculate the acceptance for a generic massless
lepton at a given energy in the squark rest frame as a function of the
lepton energy and perform a boost to the laboratory frame by means of
a simple Monte Carlo calculation. Since the squark is a spin-zero
particle, the momentum of any one of its final decay products has a
flat solid angular distribution in the squark rest frame, though of
course there are correlations between the momenta of the decay
products.  However, we find that the correlations are weak, and that
merely multiplying together acceptances for each jet or lepton
produces reasonable overall acceptances. The numerical estimate of
boosting to the laboratory frame depends on the dynamics of the squark
and gluino production process only and is stored in a grid as a
function of the squark and gluino masses and the lepton energy. We
finally multiply the generic acceptance estimate with the number of
leptons at a given energy, as obtained from the analytic calculation
of the squark-to-lepton decay.  Though the generic acceptances were
obtained for a particular choice of transverse momentum cut ($10$
GeV), they scale simply and can be re-used for different choices of
the cut.

\subsubsection{Verification}

To validate our acceptance cut estimates, we performed full
parton-level simulations, inclu\-ding spin correlations, with
\Herwig~\cite{Bahr:2008pv} for a set of mSUGRA points chosen randomly
with flat priors, restricted then to those points with spectra such
that the gluino decayed to on-shell squarks, and that the lightest
neutralino was the LSP.  The jet and missing energy acceptances from
the \Herwig\ simulations were compared to those calculated using the
methods described above, and were found to agree within $5\%$ or
better. We thus attach an uncertainty to our acceptance cut estimate
of 5\%.

\section{Numerical results}
\label{sec:numres}

The calculation of the event rates described in
section~\ref{sec:cross_sections} is available in the form of a \Cpp\
code and has been incorporated into a currently private modification
of \Fittino. It will become part of the next official release of the
{\tt Fittino} program library.  \Fittino\ is a program which attempts
to find the best fit for Lagrangian parameters of supersymmetric
models given a set of observables with uncertainties and their
covariance matrix.  One can choose to perform a fit using simulated
annealing or Markov chains (see~\cite{Bechtle:2009ty} and references
therein), and one can choose to fit to various sets of Lagrangian
parameters defined at either the TeV scale or the GUT scale.  Since
the results presented below were obtained by performing fits with
Markov chains to parameters defined at the GUT scale, we briefly
outline this process.

Firstly, \Fittino\ reads in its input file, in which all the
observables to be used are defined, with their nominal values,
uncertainties and correlations.  These observables, and the
corresponding values adopted in our analysis, are presented in
section~\ref{subsubse:Fittino_inputs}.  \Fittino\ also reads in the
parameters which should be fitted, and a set of starting values for
these parameters.  Since our aim is to show how the addition of event
rates as observables helps to reduce uncertainties on the parameters
of the fit, we began the Markov chains at the point which should be
found, rather than some other point in the parameter space.  At this
stage, our code is initialized, and loads the cross-section and
acceptance grids into memory.  Secondly, \Fittino\ begins the process
of scanning the parameter space.  At each step, it selects a random
point near its current point.  Then \SPheno~\cite{Porod:2003um} is
used to calculate the TeV-scale spectrum and branching ratios.
\Fittino\ uses this spectrum to calculate theory predictions for the
observables (either with internal code, including our extension to
rates, or by calling external programs).  These predictions are then
compared to the input values and uncertainties to calculate a
likelihood for this new point, which is then used by a Metropolis
algorithm~\cite{Metropolis:1953am} to decide whether \Fittino\ moves
to this point or rejects it and stays where it is.  Finally, the
Markov chain is analyzed to calculate a best fit for the parameters
and their uncertainties.

\subsection{Fit inputs}
\label{subsubse:Fittino_inputs}

\Fittino\ allows for the inclusion of a wide variety of low-energy and
hadron as well as lepton collider observables.  For the case study
presented here, we restrict ourselves to those ob\-ser\-va\-bles which
are expected to be measured at the LHC. We consider the minimal
supergravity SPS1a benchmark point, with parameters $M_0 = 100$~GeV,
$M_{1/2} = 250$~GeV, $A_0 = -100$~GeV, $\tan\beta = 10$ and sign$(\mu)
=+1$~\cite{Allanach:2002nj}, corresponding, \eg, to the TeV-scale
masses $m_{\tilde{g}} = 606$~GeV, $m_{\tilde{u}_L} = 559$~GeV,
$m_{\tilde{e}_L} = 177$~GeV, $m_{\tilde{\chi}_2^0} = 181$~GeV and
$m_{\tilde{\chi}_1^0} = 97.1$~GeV.  SPS1a has been studied in great
detail in experimental simulations by the ATLAS
collaboration~\cite{Weiglein:2004hn}.  Because of the comparably light
spectrum, this benchmark point provides a rich phenomenology, even at
low LHC collision energy and luminosity. Furthermore, SPS1a has been
adopted as an input for previous SUSY parameter determinations from
LHC data~\cite{Lafaye:2007vs, Bechtle:2009ty}, the results of which
may be compared to our's.

The standard set of LHC observables for SUSY mass determination
comprises the endpoints of the four invariant mass distributions that
can be constructed from the jet and the two leptons in the cascade
depicted in Fig.~\ref{fig:cascade}:\\[-6mm]
\begin{itemize}
\item[--] $m_{{\ell}{\ell}}^{\rm{max}}$, the dilepton invariant mass edge,\\[-7mm]
\item[--] $m_{q{\ell}{\ell}}^{\rm{max}}$, the jet-dilepton invariant mass edge,\\[-7mm]
\item[--] $m_{q{\ell}}^{\rm{low}}$, the jet-lepton low invariant mass edge,
  and \\[-7mm]
\item[--] $m_{q{\ell}}^{\rm{high}}$, the jet-lepton high invariant mass
  edge.\\[-5mm]
\end{itemize}
The definition of these edges and their relation to the sparticle
masses in the decay chain can be found in~\cite{Allanach:2000kt}. The
$m_{{\ell}{\ell}}^{\rm{max}}$, $m_{q{\ell}{\ell}}^{\rm{max}}$,
$m_{q{\ell}}^{\rm{low}}$ and $m_{q{\ell}}^{\rm{high}}$ endpoints have
been analyzed in detail by both the ATLAS and CMS collaborations for
different SUSY scenarios~\cite{Weiglein:2004hn, Aad:2009wy,
  Ball:2007zza, Mohr:2009jd, roth}. At the SPS1a benchmark point, they
can be measured at the LHC with a high accuracy of better than 5\%,
even at a low luminosity of 1~fb$^{-1}$~\cite{Weiglein:2004hn,
  Bechtle:2009ty}.  The estimated statistical uncertainties on the
measurements of the edges are collected in
Table~\ref{tab:input_values} for different LHC collision energies and
luminosities. We will present results for SUSY parameter fits that
involve the four standard edges (labeled group I) and the rates for
LHC data analysis at 7~TeV collision energy and 1~fb$^{-1}$ integrated
luminosity, and at 14~TeV for both 1~fb$^{-1}$ and 10~fb$^{-1}$.

\TABLE{
\setlength{\extrarowheight}{1mm}
\begin{tabular}{l d d d d}
  observable & \multicolumn{1}{c}{nominal} &
  \multicolumn{3}{c}{statistical uncertainty}\\[-1mm]
  & \multicolumn{1}{c}{value} & \multicolumn{1}{c}{for 7~TeV/$1$
    fb${}^{-1}$} & \multicolumn{1}{c}{for 14~TeV/$1$ fb${}^{-1}$} &
  \multicolumn{1}{c}{for 14~TeV/$10$ fb${}^{-1}$}\\
  \hline
  \hline
  {\bf group I} & & & & \\
  $m_{{\ell}{\ell}}^{\rm{max}}$ & 80.4 & 4.4 & 1.5 & 0.43 \\
  $m_{q{\ell}{\ell}}^{\rm{max}}$ & 452.1 & 36.0 & 12.0 & 3.6\\
  $m_{q{\ell}}^{\rm{low}}$ & 318.6 & 19.7 & 6.5 & 3.0\\
  $m_{q{\ell}}^{\rm{high}}$ & 396.0 & 13.5 & 4.5 & 3.9\\[1mm]
  \hline
  \hline
  {\bf group II} & & & & \\
  $m_{q{\ell}{\ell}}^{\rm{thr.}}$ & 215.6 & \multicolumn{1}{c}{-} &  22.8 & 4.1 \\
  $m_{T2}^{{\sq}}$ & 531.0 & \multicolumn{1}{c}{-} & 16.9 & 5.3 \\
  $m_{{\tau}{\tau}}^{\rm{max}}$ & 83.4 & \multicolumn{1}{c}{-} & 10.8 & 3.4\\
  $m_{tb}^{w}$ & 359.5 & \multicolumn{1}{c}{-} & 37.0 & 11.7\\
  $r_{{\se}{\tilde{{\tau}}}\rm{BR}}$ & 0.076 & \multicolumn{1}{c}{-} & 0.008 & 0.003\\[1mm]
  \hline
  \hline
 {\bf  group III} & & & & \\
  ${\Delta}m_{{\go}{\xone}}$ & 507.7 & \multicolumn{1}{c}{-} & \multicolumn{1}{c}{-} & 11.8\\
  $m_{({\neuino{4}}){\ell}{\ell}}^{\rm{max}}$ & 280.6 & \multicolumn{1}{c}{-}& \multicolumn{1}{c}{-} & 10.8 \\
  $m_{b{\ell}{\ell}}^{\rm{thr.}}$ & 195.9 & \multicolumn{1}{c}{-}& \multicolumn{1}{c}{-} & 17.0 \\
  $m_{h}$ & 109.6 & \multicolumn{1}{c}{-} & \multicolumn{1}{c}{-} & 1.2\\[1mm]
  \hline
  \hline
  {\bf Event rate} [fb]\rule{0mm}{8mm} & \multicolumn{2}{c}{$7$ TeV} & \multicolumn{2}{c}{$14$ TeV}\\
  & \multicolumn{1}{c}{nominal value} & \multicolumn{1}{c}{uncertainty}& \multicolumn{1}{c}{nominal value} & \multicolumn{1}{c}{uncertainty}\\
  $R_{jj{\Slashed{E}}_{T}}$ & \multicolumn{1}{c}{4.6 $\times 10^{3}$} & \multicolumn{1}{c}{9.1 $\times 10^{2}$} & \multicolumn{1}{c}{4.8 $\times 10^{4}$} & \multicolumn{1}{c}{9.5 $\times 10^{3}$}\\
  $R_{{\ell}{\ell}jj{\Slashed{E}}_{T}}$ & \multicolumn{1}{c}{1.6 $\times 10^{2}$} & \multicolumn{1}{c}{3.2 $\times 10^{1}$} & \multicolumn{1}{c}{1.5 $\times 10^{3}$} & \multicolumn{1}{c}{3.0 $\times 10^{2}$}\\[1mm]
  \hline
\end{tabular}
\caption{\label{tab:input_values} LHC observables for SPS1a used as
  inputs to the SUSY parameter fits. The nominal values for masses and
  branching ratios have been obtained with {\tt SPheno}. The
  uncertainty estimates on the observables of groups I, II and III are
  based on~\cite{Bechtle:2009ty} and have been rescaled as described
  in the main text. Note that the uncertainty estimates for 7~TeV are
  not based on a detailed experimental simulation but on a simple
  extrapolation from experimental studies at 14~TeV. The event rates
  include the NLO squark and gluino production cross sections, the
  branching ratios and the cut acceptances, as described in
  sections~\ref{sec:cross-sections} and \ref{sec:cut_acceptances},
  respectively. The overall uncertainty on the rates is assumed to be
  20\%, see section~\ref{sec:cross-sections} and the main text below.
  Dimensionful quantities are given in units of GeV for masses and
  invariant mass endpoints, or fb for event rates.}}

For 14~TeV collision energy, we also consider a set of observables
which involve third-generation particles, the lower endpoint of
$m_{q{\ell}{\ell}}$ and the stransverse mass~\cite{Lester:1999tx,
  Barr:2003rg} of the right-handed up squark:\\[-6mm]
\begin{itemize}
\item[--] $m_{q{\ell}{\ell}}^{\rm{thr.}}$, the jet-dilepton threshold invariant mass edge,\\[-7mm]
\item[--] $m_{T2}^{{\sq}}$, the squark stransverse mass,\\[-7mm]
\item[--] $m_{{\tau}{\tau}}^{\rm{max}}$, the di-tau invariant mass edge,\\[-7mm]
\item[--] $m_{tb}^{w}$, the weighted top-bottom invariant mass edge, and \\[-7mm]
\item[--] $r_{{\se}{\tilde{{\tau}}}\rm{BR}}$, the ratio of selectron- to stau-mediated $\neuino{2}$ decays.\\[-5mm]
\end{itemize}
This second group of observables (group II) is defined
in~\cite{Weiglein:2004hn, Bechtle:2009ty} and is much more challenging
experimentally than the edges of group~I. We are only aware of an
ATLAS study~\cite{Weiglein:2004hn, Bechtle:2009ty} which quantifies
the experimental accuracy to be expected at the LHC operating at
14~TeV.  In any case, also this second group of measurements may
contribute to the SUSY parameter determination at the highest LHC
energy and will be included in our analysis below.

The third group (group III), finally, is a set of observables which may
only be measurable with at least 10~fb$^{-1}$ of data at 14 TeV, and
comprises~\cite{Weiglein:2004hn, Bechtle:2009ty}\\[-6mm]
\begin{itemize}
\item[--] ${\Delta}m_{{\go}{\xone}}$, the mass difference between the gluino and the LSP,\\[-7mm]
\item[--] $m_{({\neuino{4}}){\ell}{\ell}}^{\rm{max}}$, the dilepton invariant mass edge from the decay of a ${\neuino{4}}$,\\[-7mm]
\item[--] $m_{b{\ell}{\ell}}^{\rm{thr.}}$, the $b$-tagged jet-dilepton
  threshold invariant mass edge, and \\[-7mm]
\item[--] $m_{h}$, the mass of the lightest neutral scalar Higgs
  boson.
\end{itemize}

For all SUSY parameter fits we include the two types of event rates
defined in section~\ref{sec:cross_sections} as additional
observables:\\[-6mm]
\begin{itemize}
\item[--] $R_{jj{\Slashed{E}}_{T}}$, the inclusive event rate for at
  least two hard jets with missing transverse energy, and \\[-7mm]
\item[--] $R_{{\ell}{\ell}jj{\Slashed{E}}_{T}}$, the exclusive event rate for at least two hard
  jets with missing transverse energy plus a pair of opposite-sign
  same-flavour light leptons.
\end{itemize}
The theo\-re\-ti\-cal uncertainty of our rate prediction includes 15\%
uncertainty on the NLO calculation of the squark and gluino cross
section and 5\% uncertainty on our acceptance estimate. Further
theoretical uncertainties arise from the renormalization group
running~\cite{Allanach:2003jw}, the conversion of Lagrangian
parameters into physical masses~\cite{Pierce:1996zz} and the
calculation of the branching ratios, all obtained at finite order in
perturbation theory. These effects are, however, estimated to be small
compared to the 15\% production cross section uncertainty and can
safely be neglected. For SPS1a, the statistical uncertainty on the
rate is also significantly smaller than the theory uncertainty: even
at 7~TeV collision energy with 1~fb$^{-1}$ integrated luminosity, we
expect about 150 signal events for the exclusive signature including
two opposite-sign same-flavour leptons. We thus simply assume a
conservative overall rate uncertainty of 20\% for our analysis of
SPS1a.

All observables are collected in Table~\ref{tab:input_values} together
with their estimated statistical uncertainties. The overall
experimental uncertainties that enter the fit also include a
systematic uncertainty given by the jet and lepton energy scale, which
is assumed to be 5\%~(1\%) and 0.2\%~(0.1\%), respectively, for
1~(10)~fb$^{-1}$~\cite{Weiglein:2004hn, Bechtle:2009ty}. The
uncertainties on the endpoints related to the lepton and jet energy
scale are considered 100\% correlated among different measurements.
Further systematic uncertainties for some of the observables of
groups~II and III can be found in~\cite{Bechtle:2009ty}.  The
statistical uncertainties in Table~\ref{tab:input_values} are
estimated by rescaling the uncertainties for 14~TeV collision energy
listed in~\cite{Bechtle:2009ty} by ${\sqrt{R_{N}}}$, where $R_{N}$ is
the ratio of expected numbers of events passing appropriate cuts when
going from LO at 14 TeV (on which the simulations
in~\cite{Bechtle:2009ty} are based) to NLO at 14~TeV or NLO at 7~TeV.

Our fits are obtained from Markov chains with $10^{5}$ steps. The
frequentist interpretation of these Markov chains is used to calculate
the errors of the parameters, see~\cite{Bechtle:2009ty} for details.
Including rates slows down the computation by only about 30\%; further
optimization is possible and will be implemented in the public release
of the code.

\subsection{Universal mSUGRA fits}
\label{sec:unires}

In this section and section~\ref{sec:nonunires} below we present the
main results of this paper and demonstrate that rates significantly
improve the fits of SUSY parameters at the LHC. We first discuss the
determination of the universal mSUGRA parameters $M_{0}$, $M_{1/2}$,
$\tan \beta$ and $A_{0}$ (with sign$(\mu) = +1$ fixed) for the SPS1a
benchmark scenario. We show results for 7~TeV LHC collision energy
with 1~fb$^{-1}$ luminosity, and 14~TeV with both 1~fb$^{-1}$ and
10~fb$^{-1}$.

Let us first consider LHC data taken at 7~TeV and 1~fb$^{-1}$. As
mentioned above, it is not clear which observables beyond the four
standard edges can be measured with what accuracy during such an
initial phase of LHC data analysis.  We thus address the question of
what information can be obtained from the four edges and the event
rates only. Remarkably, we find that we can constrain the universal
scalar and gaugino masses with an error of 10\% and 3\%, respectively.
The results of our fit are collected in
Table~\ref{tab:SPS1a_universal}.  Moreover, the fit reproduces the
correct value of $\tan\beta$ with a reasonable error. Note that the
tri-linear coupling $A_0$ remains essentially unconstrained. The
impact of the rate information is crucial: including the four edges of
group I alone leads to very large errors and an unreliable fit, as one
can judge from the profile likelihood contours shown in
Figure~\ref{fig:plot_M0_M12_7TeV_1_invfb}. (Note that we needed 2
million Markov chain steps in total to derive the plot showing the fit
without rates, while the fit with rates has been obtained with $10^5$
steps only.)

\TABLE{
\setlength{\extrarowheight}{1mm}
\begin{tabular}{c c c c c}
 & $M_{0}$~[GeV] & $M_{1/2}$~[GeV] & $\tan \beta$ & $A_{0}$~[GeV]\\
SPS1a & $100$ & $250$  & $10$ & $-100$\\
\hline \hline
\begin{boldmath}{\bf 7 TeV and 1 fb${}^{-1}$}\end{boldmath} & & & &\\ 
I $+$ rates & $99.0\;{}_{-9.1}^{+9.9}$ & $250.0\;{}_{-6.5}^{+8.7}$ & $10.7\;{}_{-8.8}^{+4.0}$ & $55.2\;{}_{-254}^{+1048}$\\[1mm]
\hline\hline
\begin{boldmath}{\bf 14 TeV and $1$ fb${}^{-1}$}\end{boldmath} & & & &\\ 
I $+$ rates & $99.7\;{}_{-5.7}^{+4.3}$ & $251.1\;{}_{-5.8}^{+7.5}$ & $11.2\;{}_{-5.1}^{+3.5}$ & $-50.9\;{}_{-350}^{+1233}$\\
I $+$ II, \cancel{rates} & $99.8\;{}_{-4.4}^{+3.3}$ & $249.7\;{}_{-5.2}^{+6.6}$ & $10.1\;{}_{-3.2}^{+3.8}$ & $-94.1\;{}_{-216}^{+1610}$\\
I $+$ II $+$ rates & $99.8\;{}_{-4.2}^{+3.9}$ & $251.3\;{}_{-5.0}^{+5.0}$ & $10.7\;{}_{-3.1}^{+3.1}$ & $-55.7\;{}_{-233}^{+263}$\\[1mm]
\hline\hline
\begin{boldmath}{\bf 14 TeV and $10$ fb${}^{-1}$}\end{boldmath} & & & &\\ 
I $+$ rates & $100.0\;{}_{-3.2}^{+2.9}$ & $250.7\;{}_{-3.0}^{+2.9}$ & $11.0\;{}_{-3.1}^{+2.5}$ & $-63.3\;{}_{-192}^{+165}$\\
I $+$ II, \cancel{rates} & $100.1\;{}_{-1.9}^{+1.7}$ & $250.4\;{}_{-1.7}^{+1.2}$ & $10.1\;{}_{-1.0}^{+1.1}$ & $-89.8\;{}_{-80.3}^{+70.4}$\\
I $+$ II $+$ rates & $100.3\;{}_{-1.9}^{+1.6}$ & $250.4\;{}_{-1.6}^{+1.4}$ & $10.2\;{}_{-1.0}^{+1.2}$ & $-96.5\;{}_{-68.5}^{+86.3}$\\
I $+$ II $+$ III, \cancel{rates} & $100.2\;{}_{-1.6}^{+1.4}$ & $250.3\;{}_{-1.4}^{+1.1}$ & $10.1\;{}_{-0.8}^{+0.8}$ & $-94.6\;{}_{-55.0}^{+48.2}$\\
I $+$ II $+$ III $+$ rates & $100.1\;{}_{-1.5}^{+1.6}$ & $250.3\;{}_{-1.4}^{+1.1}$ & $10.3\;{}_{-1.0}^{+0.7}$ & $-90.3\;{}_{-57.7}^{+52.1}$\\[1mm]
\hline\hline
\end{tabular}
\caption{\label{tab:SPS1a_universal} Fits to universal mSUGRA
  parameters for SPS1a, with (``+rates'') and without
  (``\cancel{rates}'') event rates as an observable. The symbols 
  I, II and III refer to the inclusion of the groups of previously considered observables (mainly edges) defined in the main text. }}

\FIGURE{
\begin{tabular}{ c c }
I, \cancel{rates} & I $+$ rates\\
  \includegraphics[width=0.5\textwidth,
  angle=0]{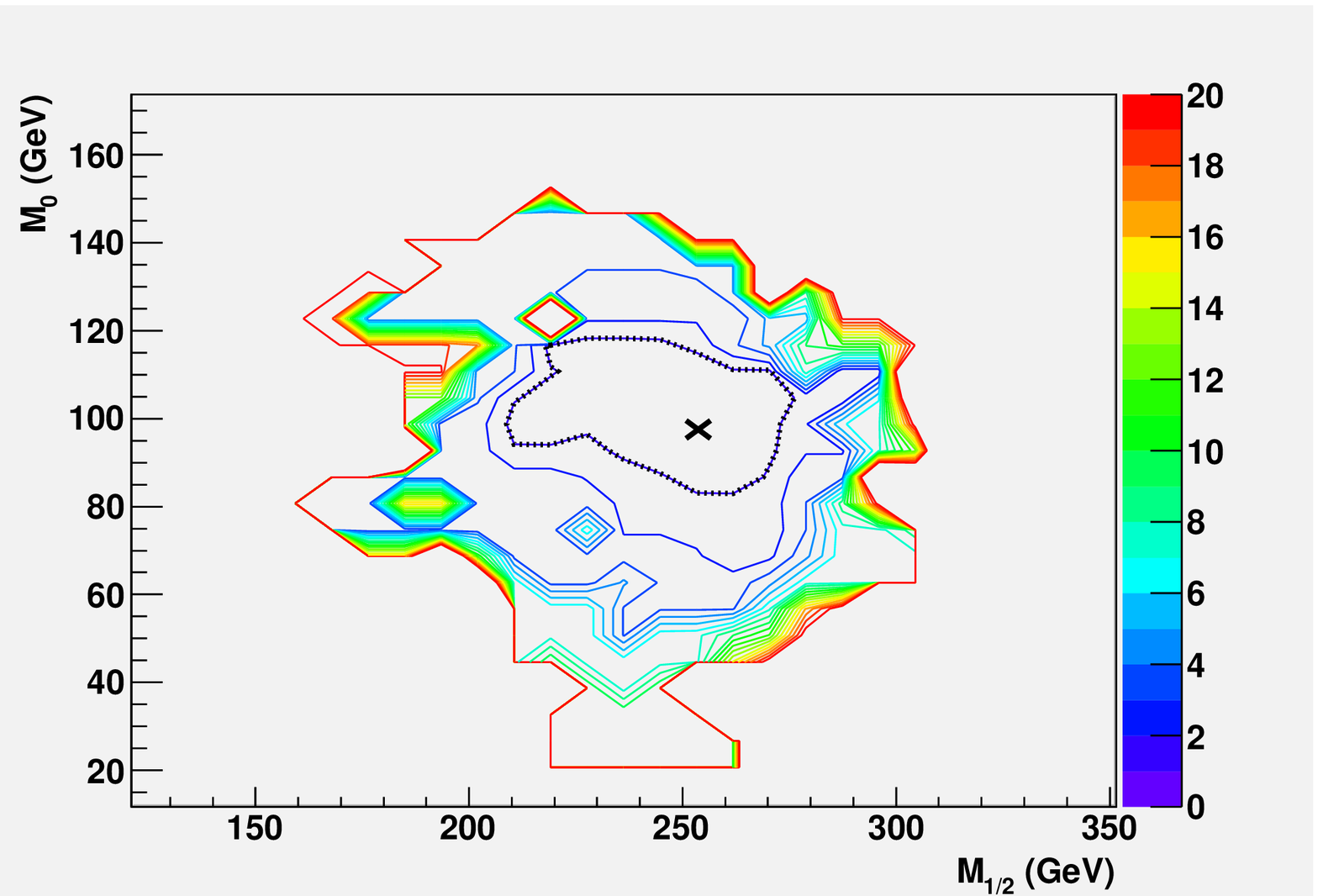}
&
  \includegraphics[width=0.5\textwidth,
  angle=0]{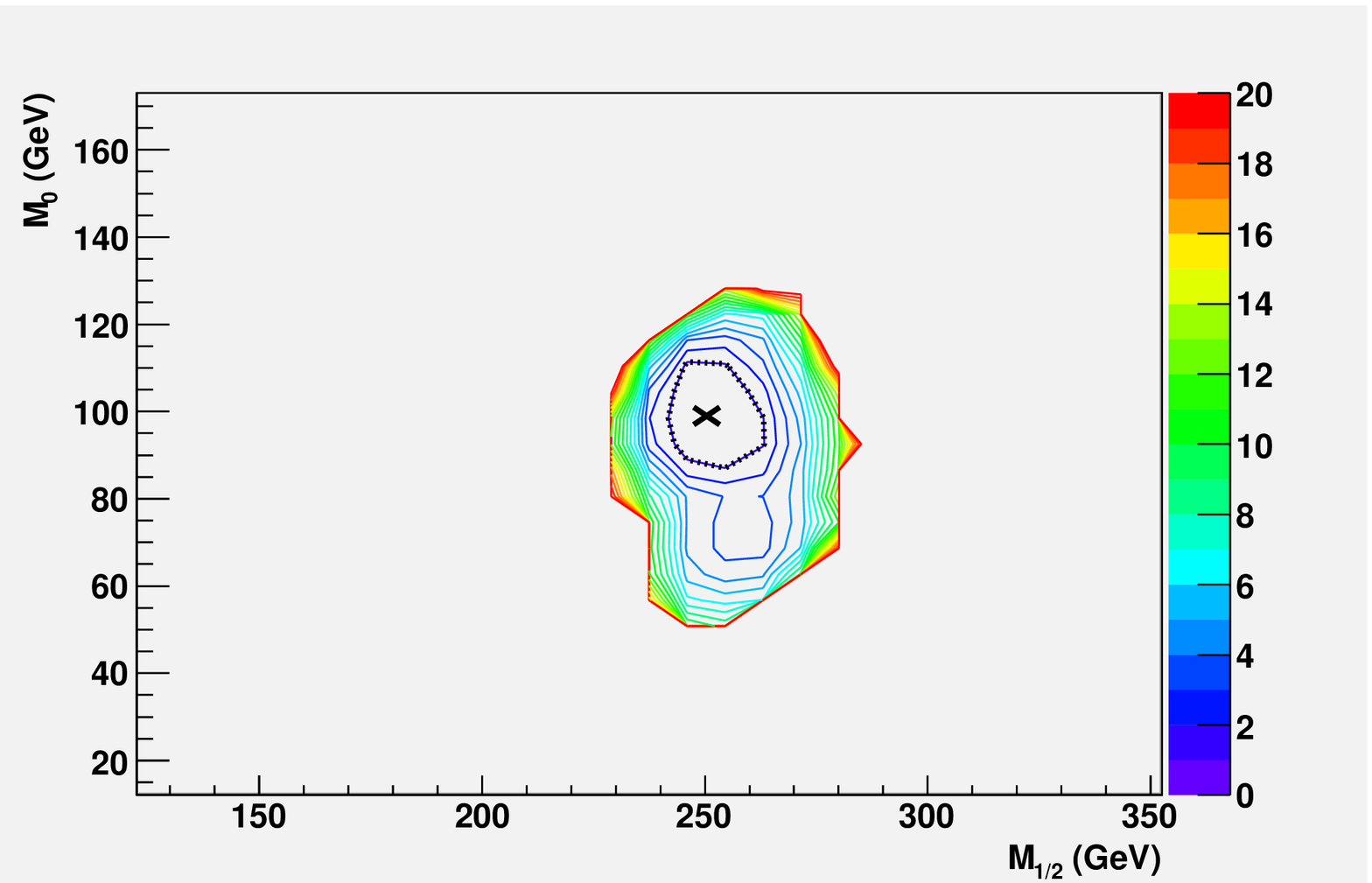}
\\
\end{tabular}
\caption{\label{fig:plot_M0_M12_7TeV_1_invfb} $\Delta\chi^2 = -2
  \ln{\cal L} + 2\ln{\cal L}_{\rm max}$ contours showing $M_{0}$
  against $M_{1/2}$ for 7 TeV/1 fb${}^{-1}$ data. Fits are based on
  the four standard edges of group~I without rates (``I,
  \cancel{rates}'', left) and with rates (``I + rates'', right).
  ${\cal L}$ is the two-dimensional profile likelihood and ${\cal
    L}_{\rm max}$ the global maximum of the likelihood. The black
  dotted contours represent $\Delta\chi^2 = 1$ contours. See
  \cite{Bechtle:2009ty} for more details.}}

\FIGURE{
\begin{tabular}{ c c }
 & I $+$ rates\\
&
  \includegraphics[width=0.5\textwidth,
  angle=0]{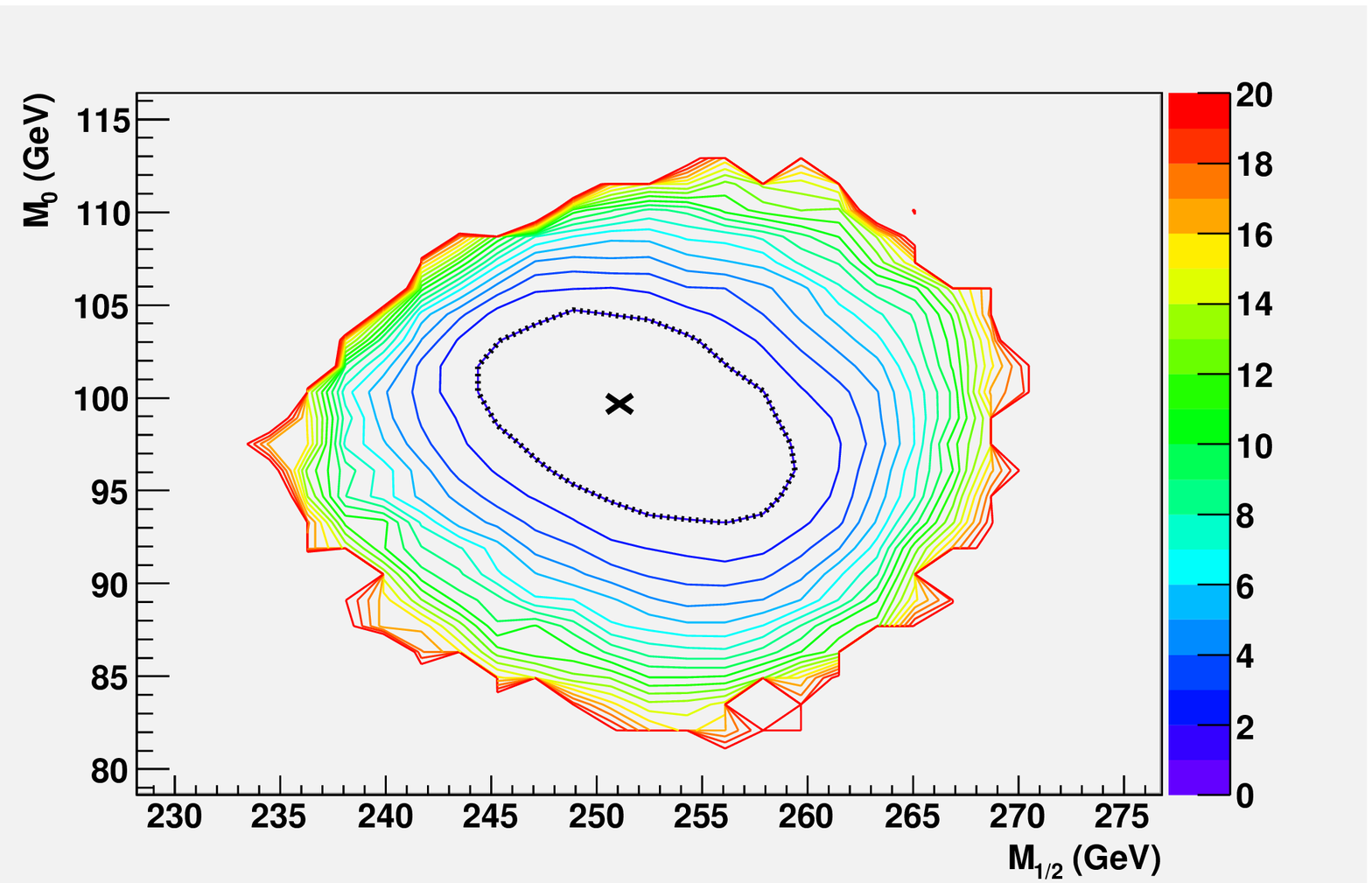}
\\
 & \\
I $+$ II, \cancel{rates} & I $+$ II $+$ rates\\
  \includegraphics[width=0.5\textwidth,
  angle=0]{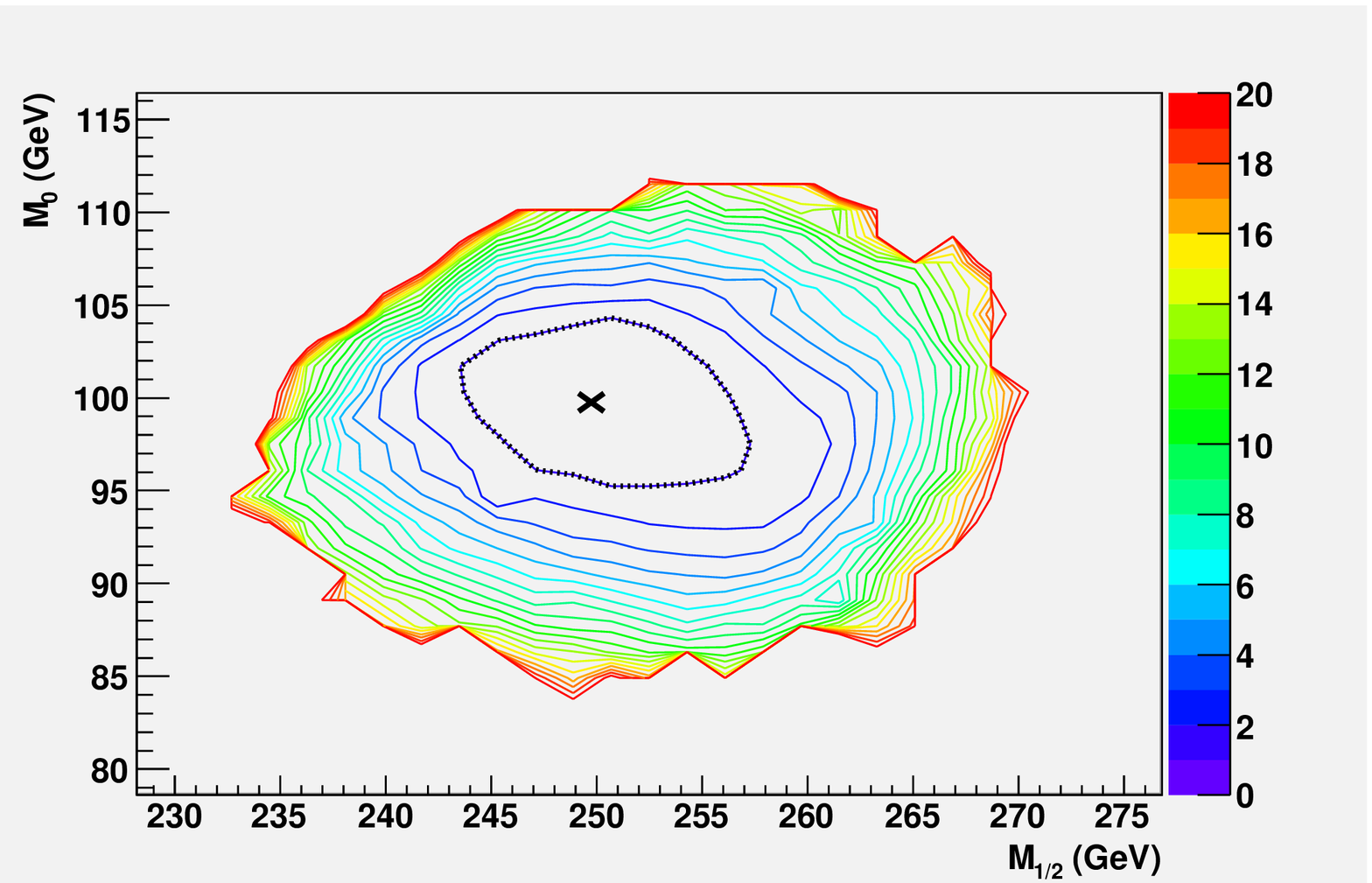}
&
  \includegraphics[width=0.5\textwidth,
  angle=0]{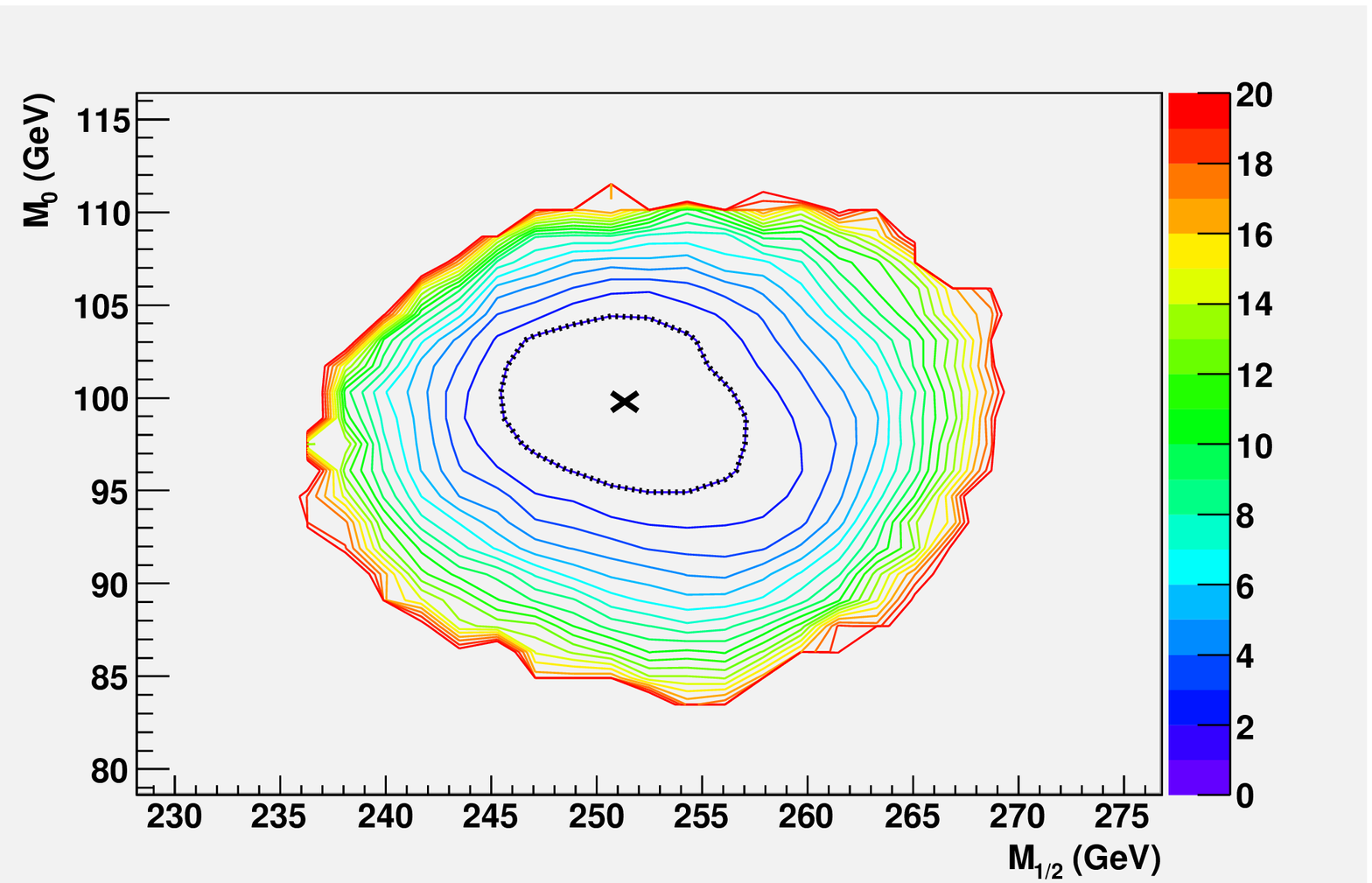}
\\
\end{tabular}
\caption{\label{fig:plot_M0_M12_14_TeV_1_invfb} $\Delta\chi^2$
  contours showing $M_{0}$ against $M_{1/2}$ for 14~TeV/1~fb${}^{-1}$
  data. Fits are based on the four standard edges of group~I with
  rates (upper right), and on the observables of groups~I and II with
  (lower right) and without rates (lower left).}}

\FIGURE{
\begin{tabular}{ c c }
 & I $+$ rates\\
&
  \includegraphics[width=0.5\textwidth,
  angle=0]{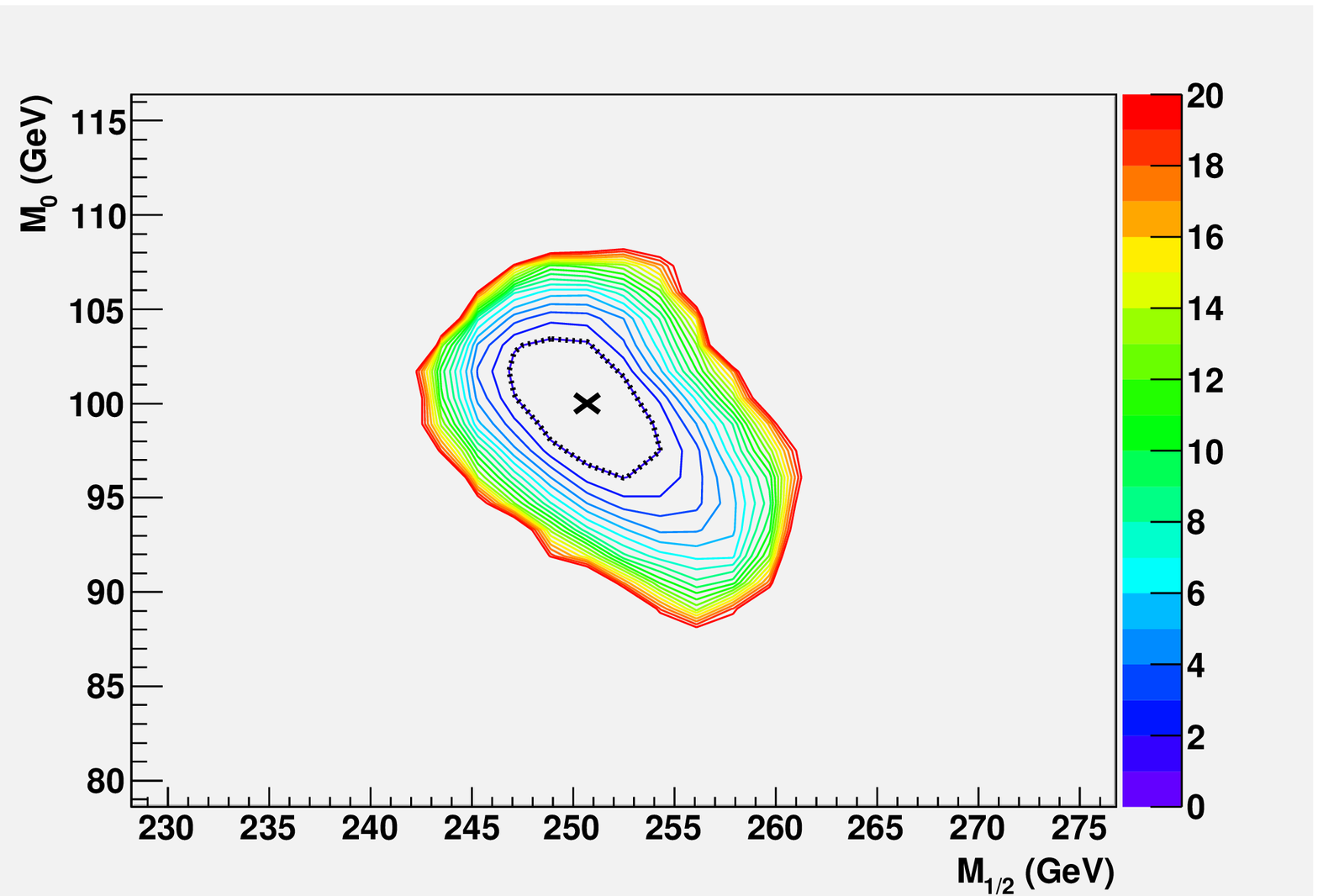}
\\
 & \\
I $+$ II, \cancel{rates} & I $+$ II $+$ rates\\
  \includegraphics[width=0.5\textwidth,
  angle=0]{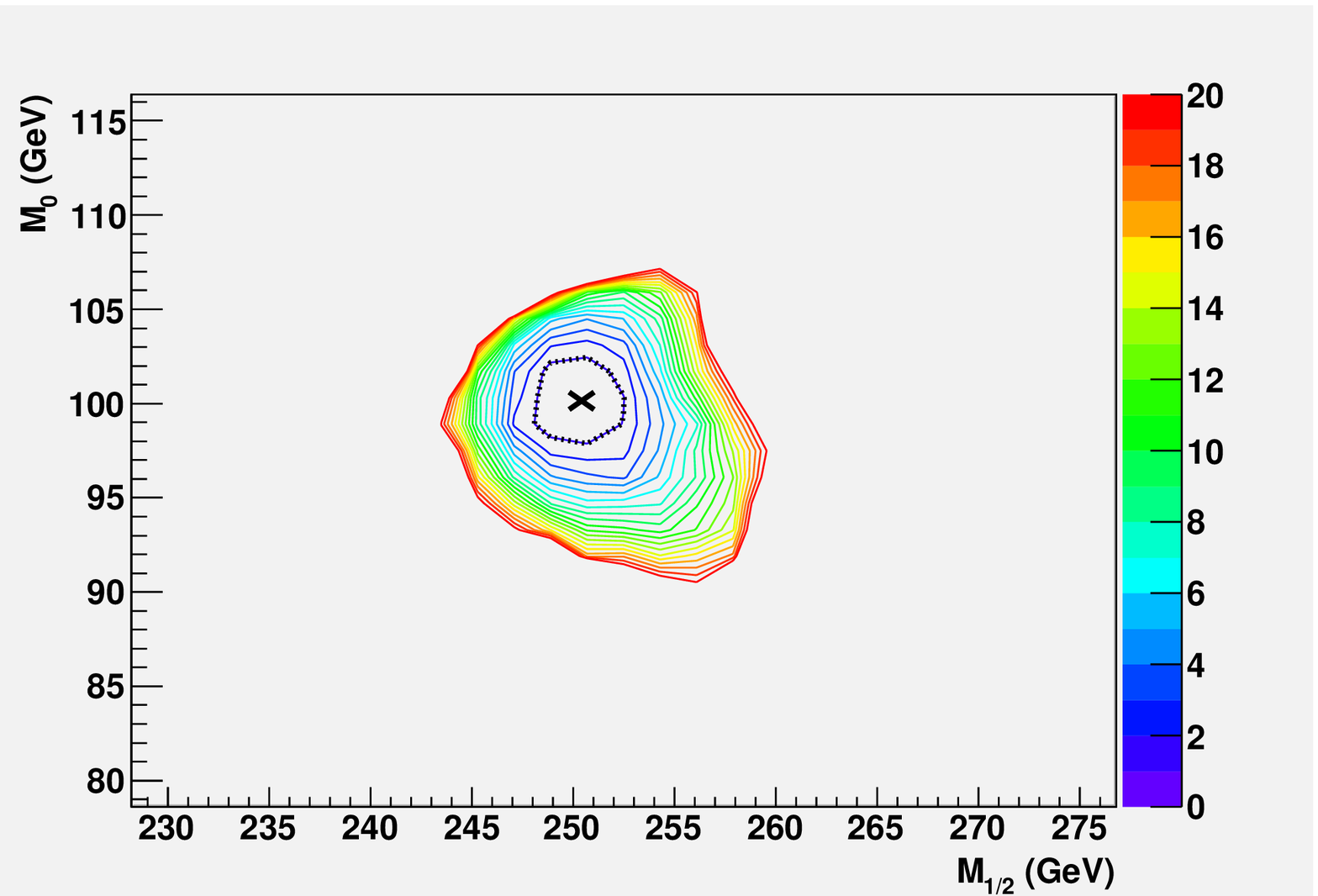}
&
  \includegraphics[width=0.5\textwidth,
  angle=0]{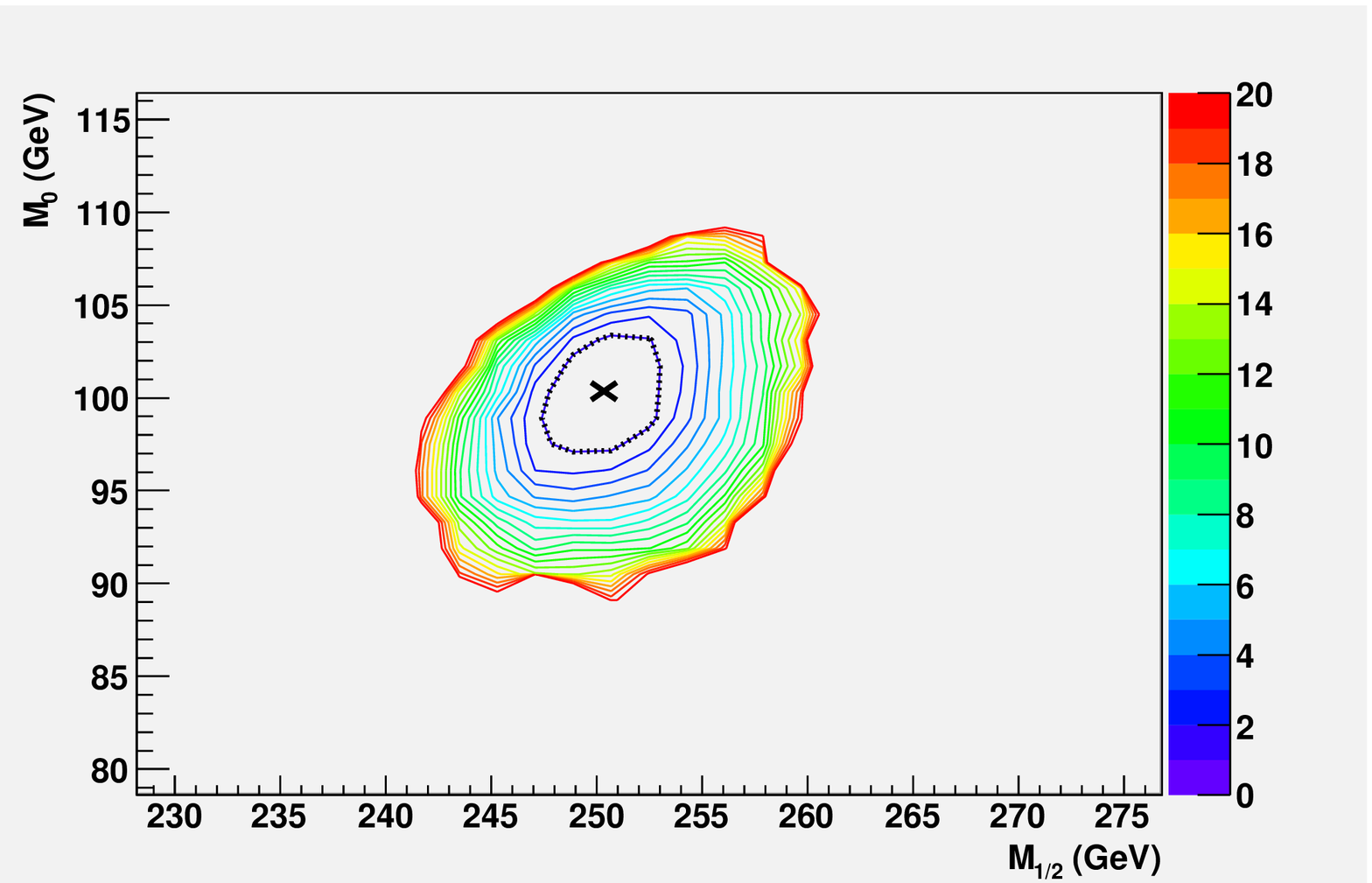}
\\
 & \\
I $+$ II $+$ III, \cancel{rates} & I $+$ II $+$ III $+$ rates\\
  \includegraphics[width=0.5\textwidth,
  angle=0]{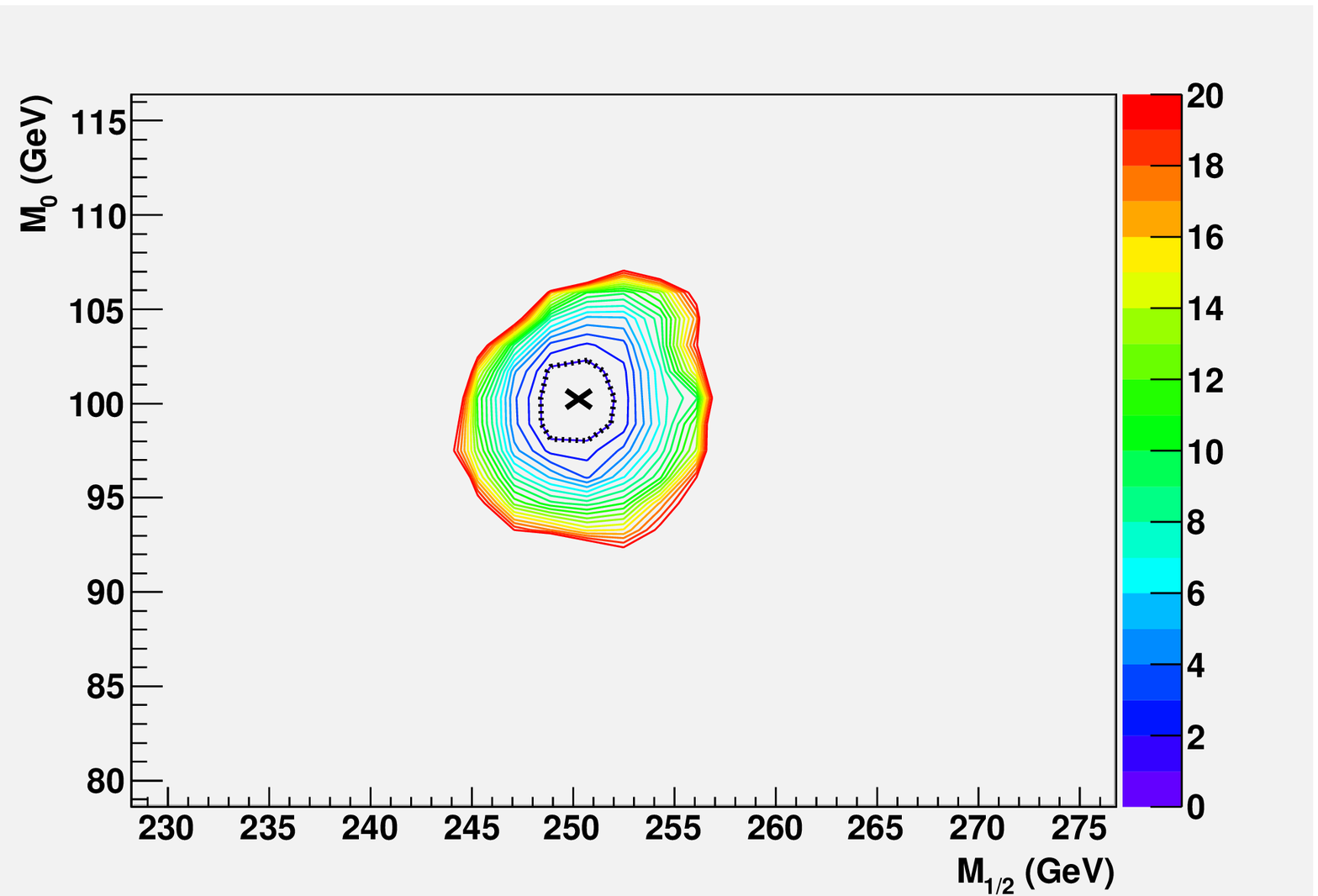}
&
  \includegraphics[width=0.5\textwidth,
  angle=0]{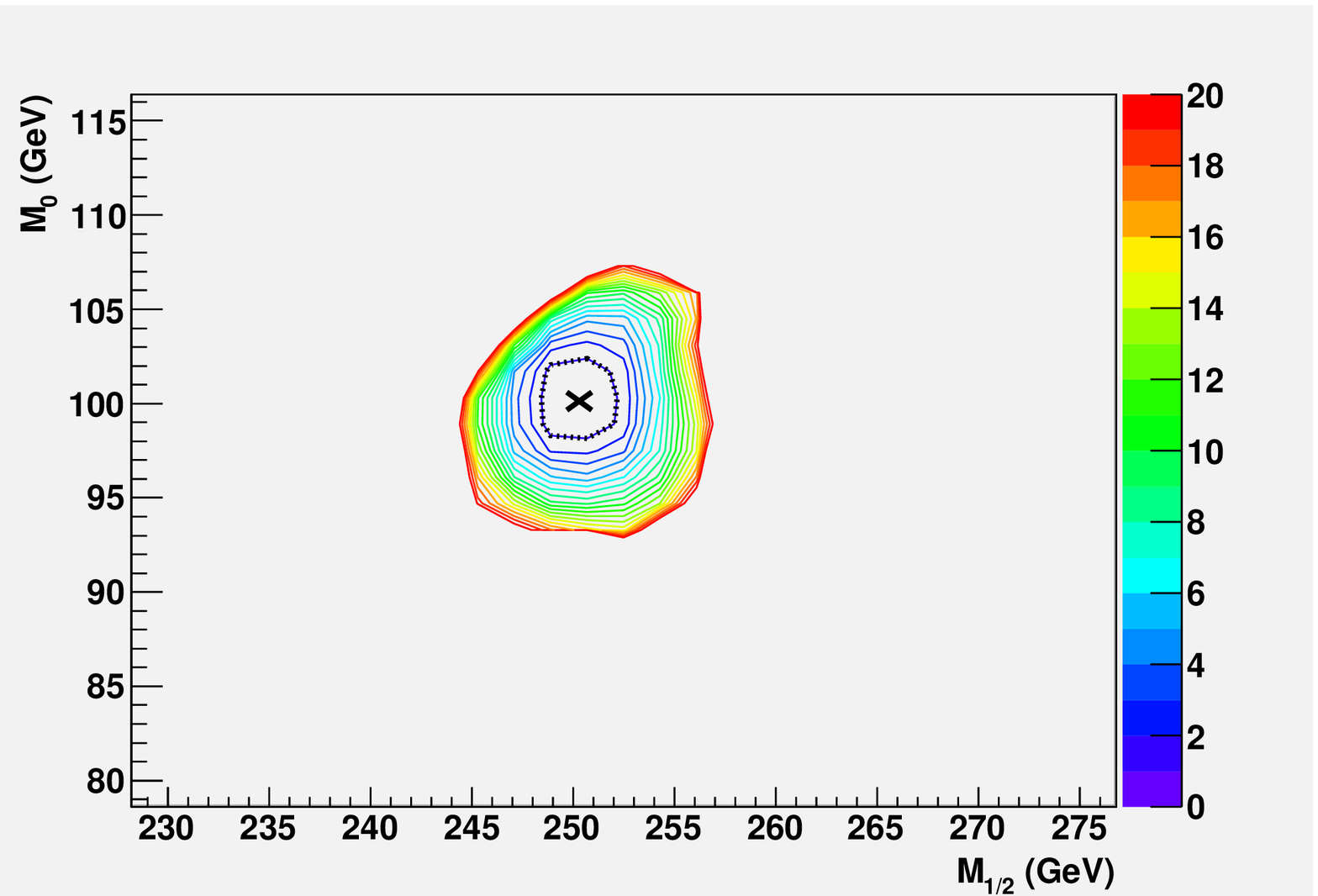}
\\
\end{tabular}
\caption{\label{fig:plot_M0_M12_14_TeV_10_invfb} $\Delta\chi^2$
  contours showing $M_{0}$ against $M_{1/2}$ for 14~TeV/10~fb${}^{-1}$
  data. Fits are based on the four standard edges of group~I with
  rates (upper right), on the observables of groups~I and II with
  (middle right) and without rates (middle left), and on observables
  of groups~I, II, and III with (lower right) and without rates (lower
  left).}}

At 14~TeV collision energy and 1~fb$^{-1}$ luminosity, the statistical
uncertainty on the measurement of edges is considerably reduced, see
Table~\ref{tab:input_values}. We thus obtain a better determination of
the mSUGRA parameters with rates and the basic edges of group~I, in
particular for the common scalar mass $M_0$. Note that the errors on
$\tan\beta$ and $M_{1/2}$ in the fit ``I + rates'' in
Table~\ref{tab:SPS1a_universal} would be twice as large had we based
our rate estimates on LO cross sections with the corresponding 100\%
uncertainty. Adding the observables of group~II, which involve
information on third-generation particles, the lower endpoint of
$m_{q{\ell}{\ell}}$ and the stransverse mass, does not lead to a
significant improvement.  The rate information together with the
standard edges thus seems sufficient to determine the parameters of an
mSUGRA fit to SPS1a-type SUSY scenarios in the initial phase of a high
energy LHC run. Our results for 14~TeV and 1~fb$^{-1}$ without rates,
presented in Table~\ref{tab:SPS1a_universal} and
Figure~\ref{fig:plot_M0_M12_14_TeV_1_invfb}, are consistent with the
results presented in Ref.~\cite{Bechtle:2009ty}, specifically Table~15
thereof.

At 14 TeV and 10~fb$^{-1}$ also the systematical errors from lepton
and jet energy scales are expected to improve, leading to a further
reduction of the error on edges. Fitting the four edges of group~I and the 
rates leads to a very accurate determination of the mSUGRA parameters 
$M_0$ and $M_{1/2}$ and of $\tan\beta$. At high energy and high luminosity, 
the observables of group~III may become accessible. 
With all 13 observables of groups~I, II and III included, the fit to
the SPS1a mSUGRA scenario is so well constrained that adding rates
does not lead to a significant improvement, see
Figure~\ref{fig:plot_M0_M12_14_TeV_10_invfb}. Our results are consistent with those
presented in Table~16 of~\cite{Bechtle:2009ty}.

\subsection{Non-universal mSUGRA fits}
\label{sec:nonunires}

Let us now briefly discuss a more general class of fits where we
attempt to determine the GUT-scale gaugino mass parameters $M_1$,
$M_2$ and $M_3$ as fit parameters individually, instead of a common
mass $M_{1/2}$. Models without gaugino mass unification at the GUT
scale can lead to very different phenomenology, see
\eg~\cite{Dreiner:2009ic, Dreiner:2009er}, and have, to our knowledge,
so far not been considered in SUSY parameter fits of LHC data alone.
As in section~\ref{sec:unires}, we use the SPS1a observables collected
in Table~\ref{tab:input_values} as input.  Even though SPS1a is
defined by $M_1 = M_2 = M_3 = 250$~GeV at the GUT scale, our results
presented below demonstrate the importance of rates for the
determination of individual gaugino mass parameters, crucial for the
analysis of more general, non-universal, models.

Analogous to what we observe for the fit of universal mSUGRA
parameters, we find that a measurement of the four standard edges of
group~I alone is not sufficient to reliably determine the SUSY
parameters of non-universal models in the initial phase of LHC at
7~TeV and with 1~fb$^{-1}$.  This is born out by the profile
likelihood contours shown in Figure~\ref{fig:plot_M0_M3_7_TeV_1_invfb}
(left). Adding rates, however, we obtain a reasonable fit of the
larger parameter space including $M_1$, $M_2$ and $M_3$, as shown in
Table~\ref{tab:SPS1a_nonuniversal} and
Figure~\ref{fig:plot_M0_M3_7_TeV_1_invfb} (right).  We find an accuracy of
about 25\% on $M_1$, $M_2$ and 5\% on $M_3$, see
Table~\ref{tab:SPS1a_nonuniversal}. The high accuracy on $M_3$ is a
result of the sensitivity of the production cross sections to the
gluino mass, \cf~Figure~\ref{fig:NLO_cross-sections_14_TeV}.

\TABLE{
\setlength{\extrarowheight}{1mm}
\begin{tabular}{ccccccc}
 & $M_{0}$~[GeV] & $M_{1}$~[GeV]  & $M_{2}$~[GeV]  & $M_{3}$~[GeV]  & $\tan \beta$ & $A_{0}$~[GeV] \\
SPS1a & $100$ & $250$ & $250$ & $250$ & $10$ & $-100$\\
\hline\hline
\begin{boldmath}{\bf 7 TeV and 1 fb${}^{-1}$}\end{boldmath} & & & &\\ 
I $+$ rates & 91.1 ${}_{-36.1}^{+27.3}$ & 236.5 ${}_{-57.9}^{+67.1}$ & 242.6${}_{-33.7}^{+51.6}$ & 251.0${}_{-8.5}^{+9.5}$& 
10.5 ${}_{-7.3}^{+7.4}$ & $-6.0$  ${}_{-582}^{+1088}$\\[1mm]
\hline\hline
\begin{boldmath}{\bf 14 TeV and $1$ fb${}^{-1}$}\end{boldmath} & & & & & & \\ 
I $+$ rates & 98.5  ${}_{-18.4}^{+16.5}$ & 245.8 ${}_{-40.7}^{+55.7}$ & 244.2 ${}_{-19.4}^{+42.1}$ & 250.3 ${}_{-7.0}^{+11.1}$ & 6.2 ${}_{-2.0}^{+11.0}$ & $-389.9$ ${}_{-169}^{+2195}$\\
I $+$ II, \cancel{rates} & 102.7  ${}_{-21.4}^{+9.4}$ & 258.0  ${}_{-51.1}^{+32.5}$ & 255.4  ${}_{-41.7}^{+43.6}$ & 251.4  ${}_{-12.2}^{+9.9}$ & 10.3  ${}_{-3.0}^{+5.9}$ & $-102.0$  ${}_{-186}^{+1377}$\\
I $+$ II $+$ rates & 98.6  ${}_{-11.2}^{+12.6}$ & 249.6  ${}_{-24.7}^{+31.7}$ & 248.7  ${}_{-15.5}^{+24.9}$ & 252.1  ${}_{-7.1}^{+6.0}$ & 9.5  ${}_{-2.7}^{+6.6}$ & $-127.5$  ${}_{-204}^{+790}$\\[1mm]
\hline\hline
\begin{boldmath}{\bf 14 TeV and $10$ fb${}^{-1}$}\end{boldmath} & & & & & & \\ 
I $+$ rates & 99.2 ${}_{-9.3}^{+13.3}$ & 253.7 ${}_{-30.5}^{+37.1}$ & 256.9 ${}_{-23.9}^{+20.2}$ & 250.9 ${}_{-5.7}^{+8.6}$ & 14.3 ${}_{-9.7}^{+1.7}$ & $186.6$ ${}_{-761}^{+239}$\\
I $+$ II, \cancel{rates} & 98.4.0 ${}_{-4.1}^{+7.9}$ & 246.8 ${}_{-13.0}^{+22.5}$ & 248.0 ${}_{-6.8}^{+12.6}$ & 249.2 ${}_{-3.2}^{+5.2}$ & 9.6 ${}_{-0.8}^{+1.7}$ & $-117.5$ ${}_{-45.5}^{+83.1}$\\
I $+$ II $+$ rates & 102.0 ${}_{-5.7}^{+2.5}$ & 258.3 ${}_{-20.0}^{+9.2}$ & 254.3 ${}_{-10.2}^{+6.0}$ & 251.9 ${}_{-5.4}^{+2.1}$ & 9.9 ${}_{-1.0}^{+1.4}$ & $-124.7$ ${}_{-63.6}^{+101}$\\
I $+$ II $+$ III, \cancel{rates} & 99.1 ${}_{-4.6}^{+5.0}$ & 245.9 ${}_{-8.9}^{+19.5}$ & 248.4 ${}_{-5.2}^{+10.3}$ & 248.7 ${}_{-2.2}^{+5.2}$ & 9.9 ${}_{-0.7}^{+0.9}$ & $-98.6$ ${}_{-48.0}^{+41.7}$\\
I $+$ II $+$ III $+$ rates & 99.1 ${}_{-3.5}^{+6.4}$ & 251.3 ${}_{-11.2}^{+17.5}$ & 250.4 ${}_{-5.1}^{+9.9}$ & 251.2 ${}_{-4.6}^{+3.0}$ & 9.9 ${}_{-0.7}^{+1.1}$ & $-101.0$ ${}_{-50.5}^{+47.9}$\\[1mm]
\hline\hline
\end{tabular}
\caption{\label{tab:SPS1a_nonuniversal} Fits to mSUGRA parameters with
  non-universal gaugino masses for SPS1a, with (``+rates'') and without
  (``\cancel{rates}'') event rates
  as an observable.  I, II and III refer to the inclusion of the
  groups of previously considered observables (mainly edges) defined
  in the main text. }}

\FIGURE{
\begin{tabular}{ c c }
I, \cancel{rates} & I $+$ rates\\
  \includegraphics[width=0.5\textwidth,
  angle=0]{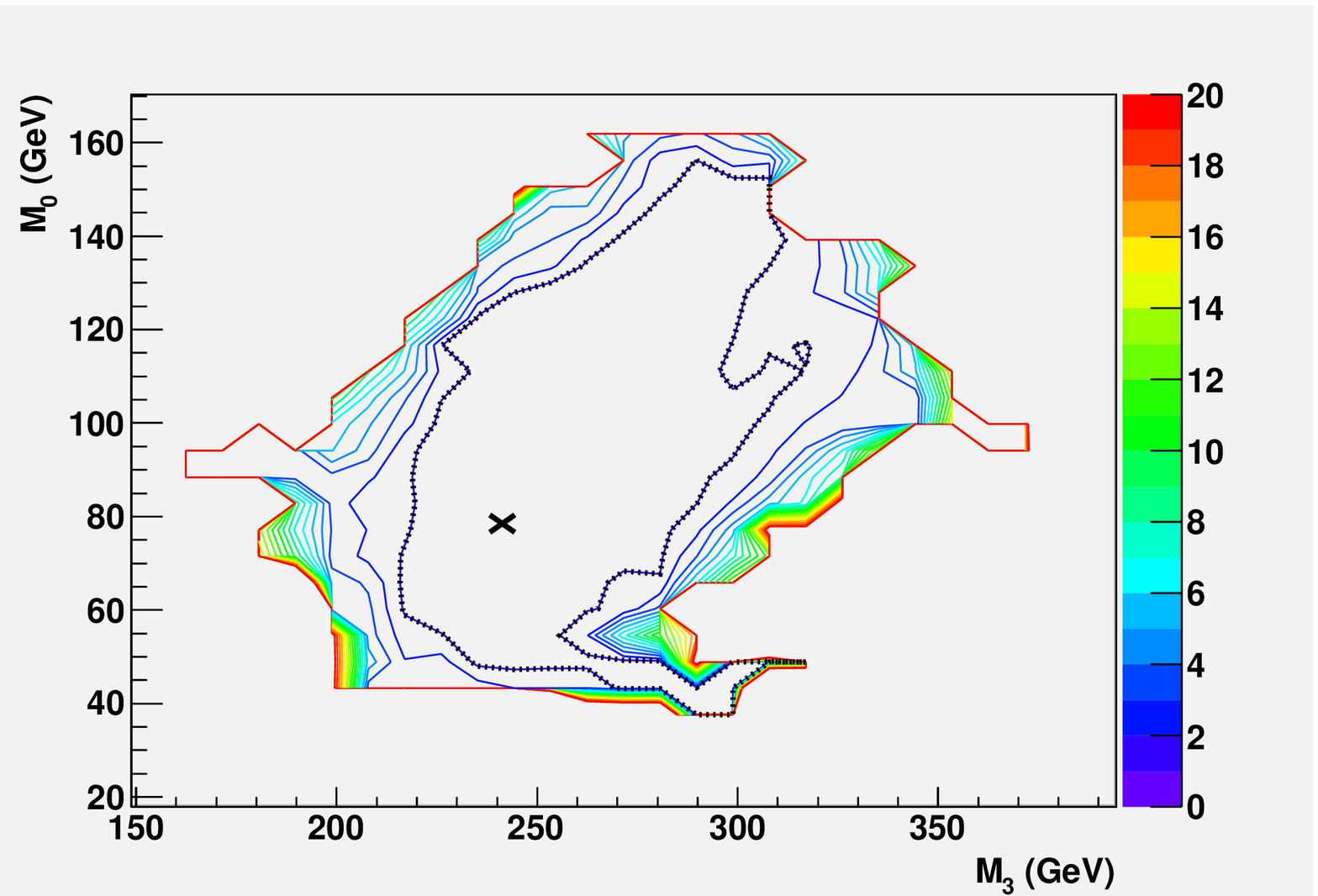}
&
  \includegraphics[width=0.5\textwidth,
  angle=0]{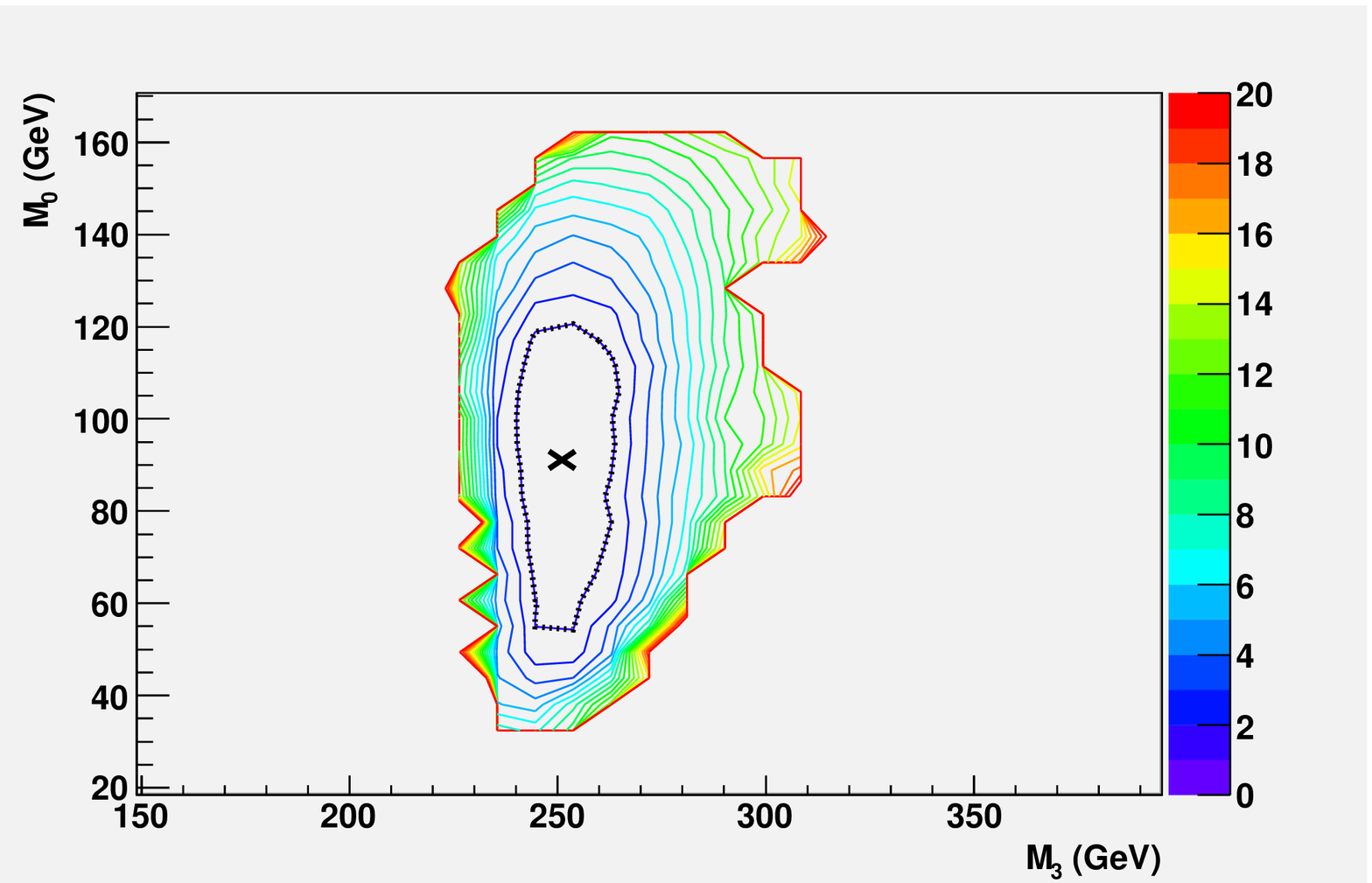}
\\
\end{tabular}
\caption{\label{fig:plot_M0_M3_7_TeV_1_invfb} $\Delta\chi^2$ contours
  showing $M_{0}$ against $M_{3}$ for 7~TeV/1~fb${}^{-1}$ data. Fits
  are based on the four standard edges of group~I without (left) and
  with rates (right).}}

\FIGURE{
\begin{tabular}{ c c }
 & I $+$ rates\\
&
  \includegraphics[width=0.5\textwidth,
  angle=0]{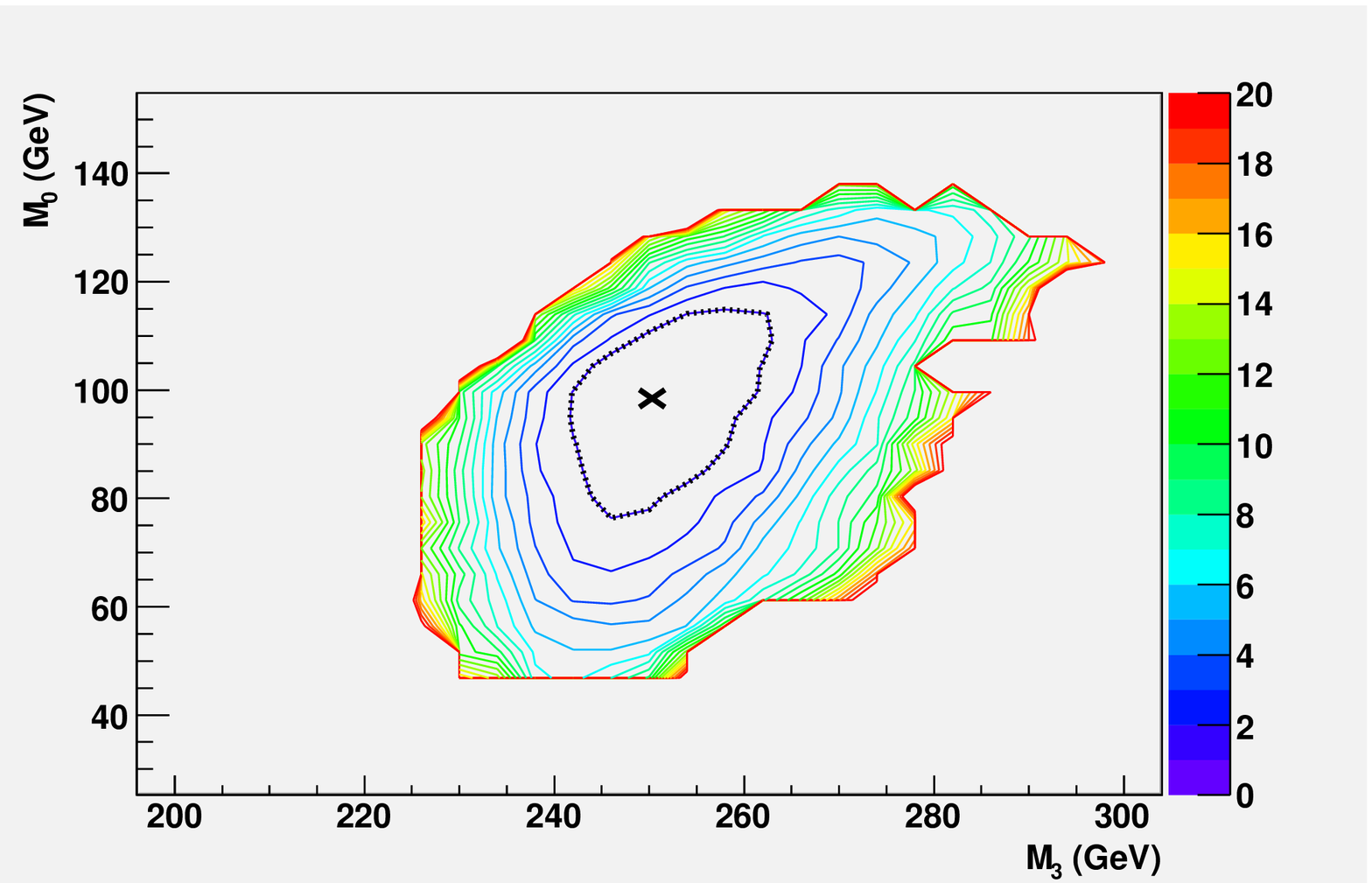}
\\
 & \\
I $+$ II, \cancel{rates} & I $+$ II $+$ rates\\
  \includegraphics[width=0.5\textwidth,
  angle=0]{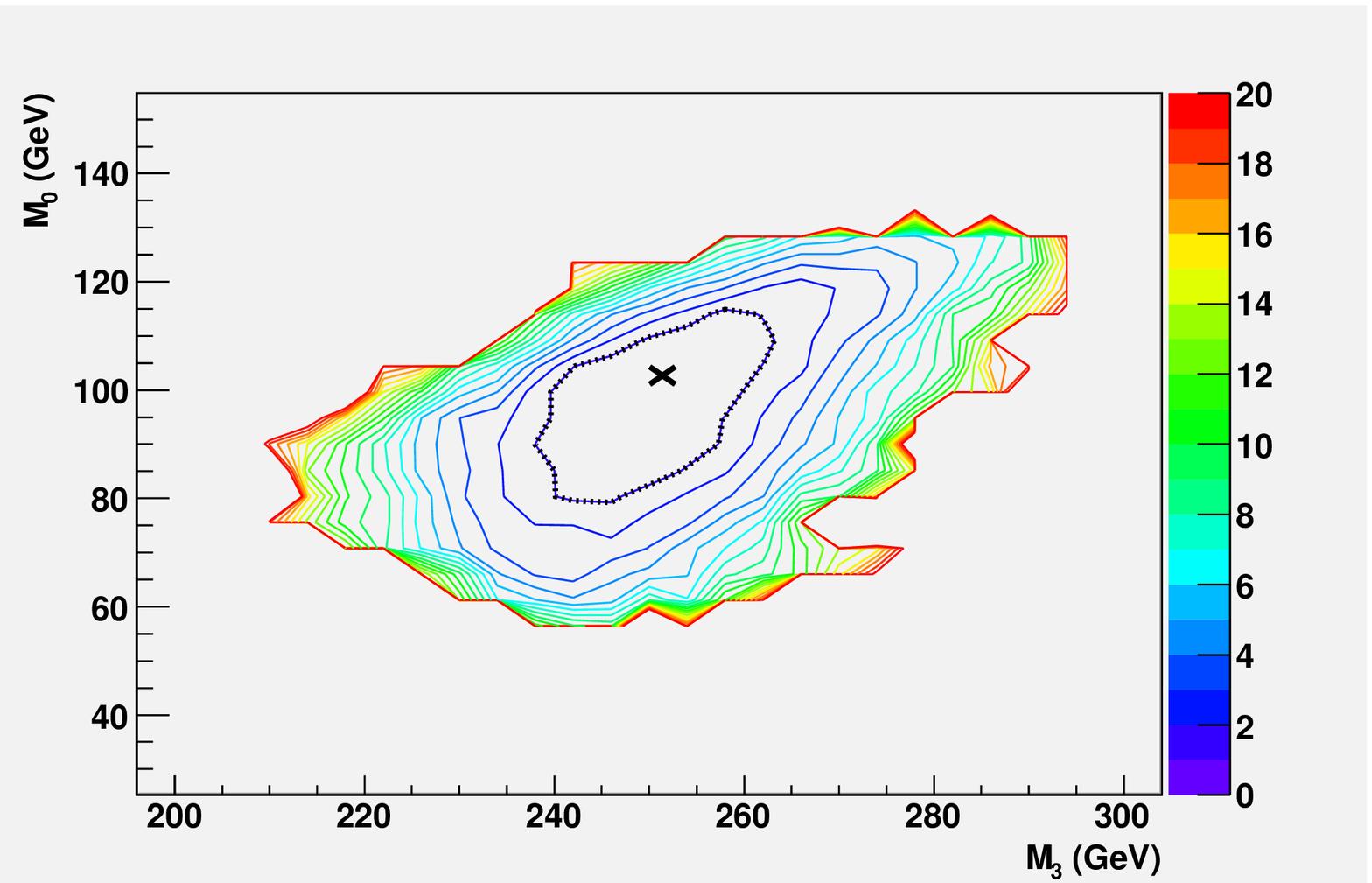}
&
  \includegraphics[width=0.5\textwidth,
  angle=0]{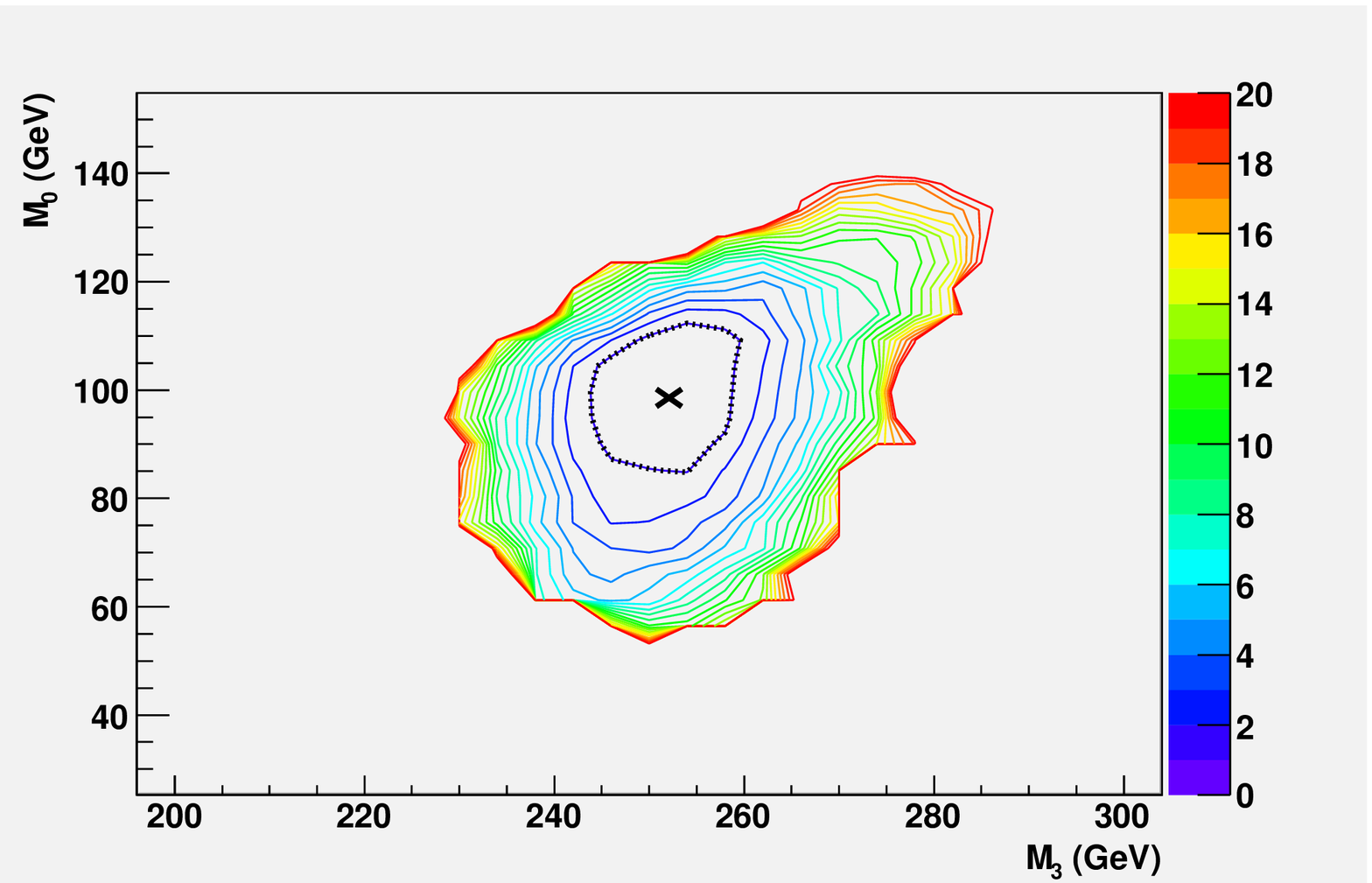}
\\
\end{tabular}
\caption{\label{fig:plot_M0_M3_14_TeV_1_invfb} $\Delta\chi^2$ contours
  showing $M_{0}$ against $M_{3}$ for 14~TeV/1~fb${}^{-1}$ data. Fits
  are based on the four standard edges of group~I with rates (upper
  right), and on the observables of groups~I and II with (lower right)
  and without rates (lower left).}}

\FIGURE{
\begin{tabular}{ c c }
 & I $+$ rates\\
&
  \includegraphics[width=0.5\textwidth,
  angle=0]{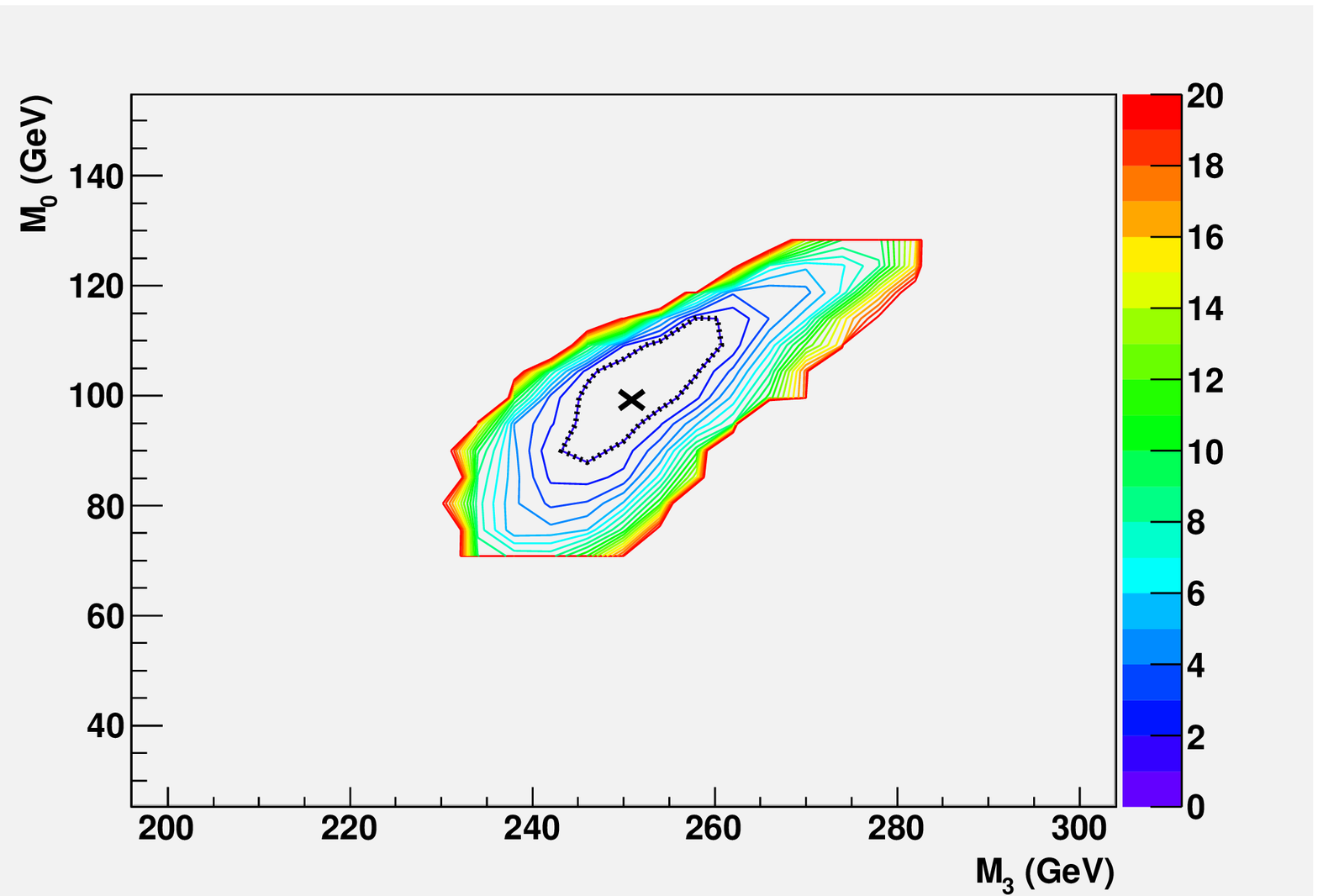}
\\
 & \\
I $+$ II, \cancel{rates} & I $+$ II $+$ rates\\
  \includegraphics[width=0.5\textwidth,
  angle=0]{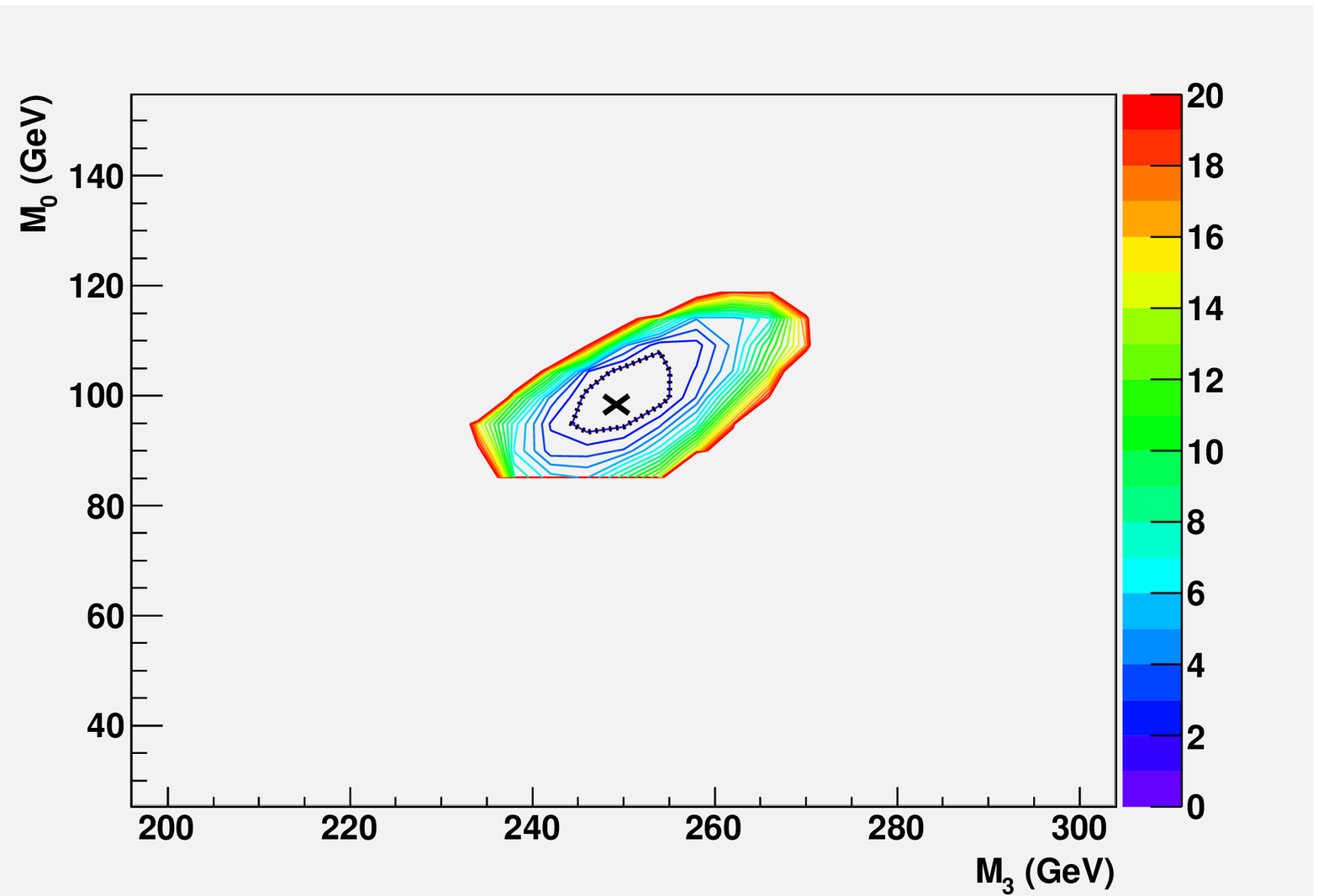}
&
  \includegraphics[width=0.5\textwidth,
  angle=0]{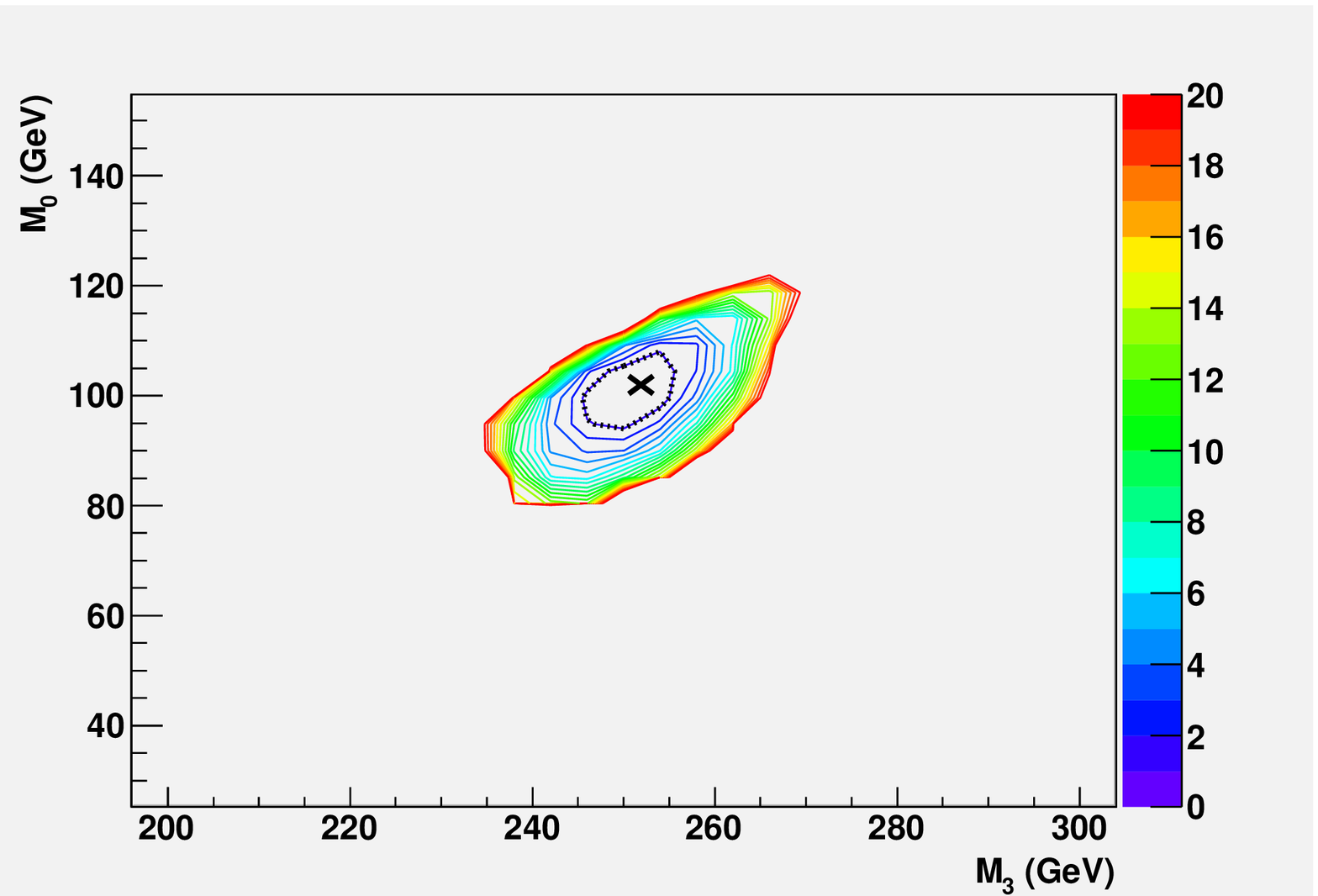}
\\
 & \\
I $+$ II $+$ III, \cancel{rates} & I $+$ II $+$ III $+$ rates\\
  \includegraphics[width=0.5\textwidth,
  angle=0]{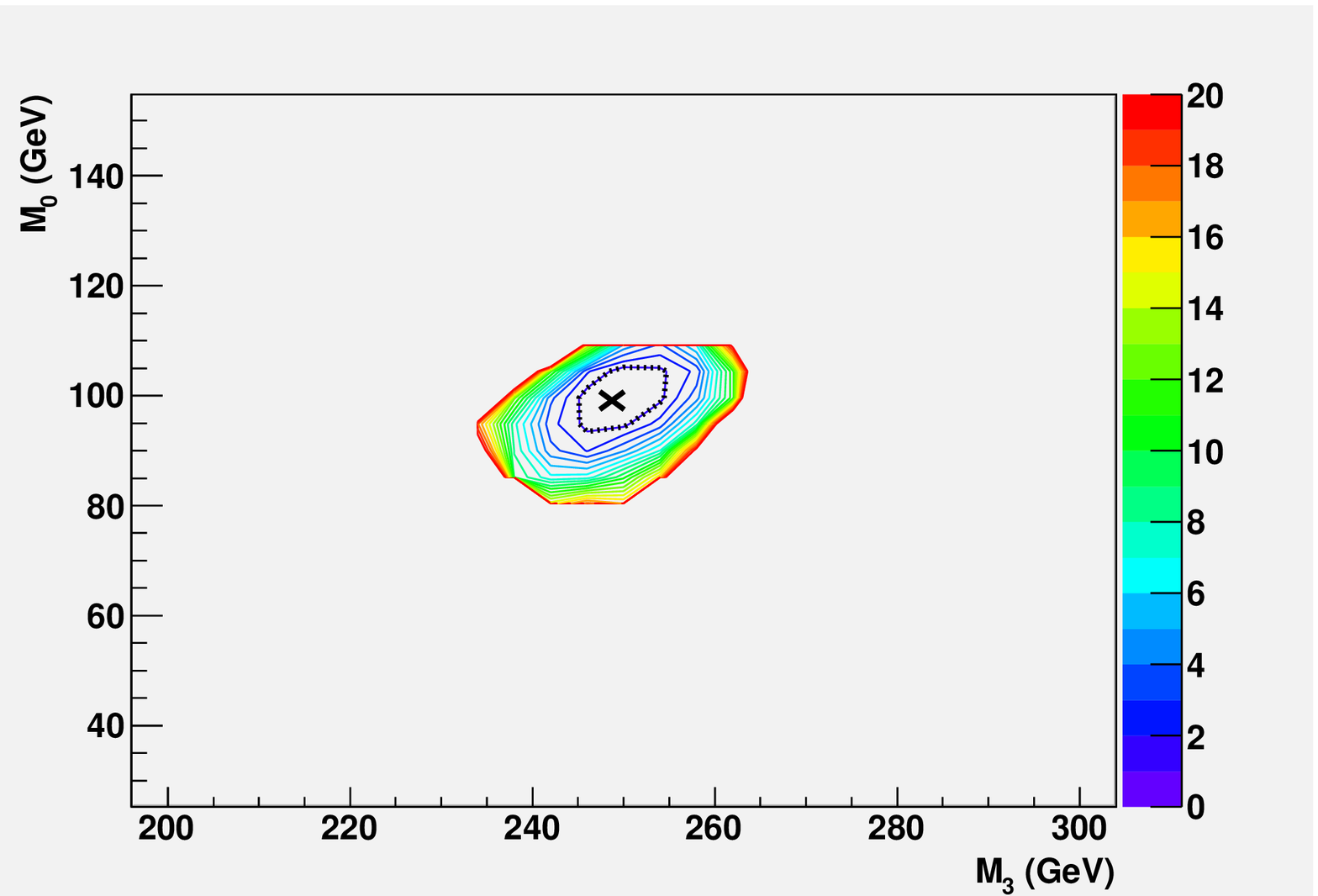}
&
  \includegraphics[width=0.5\textwidth,
  angle=0]{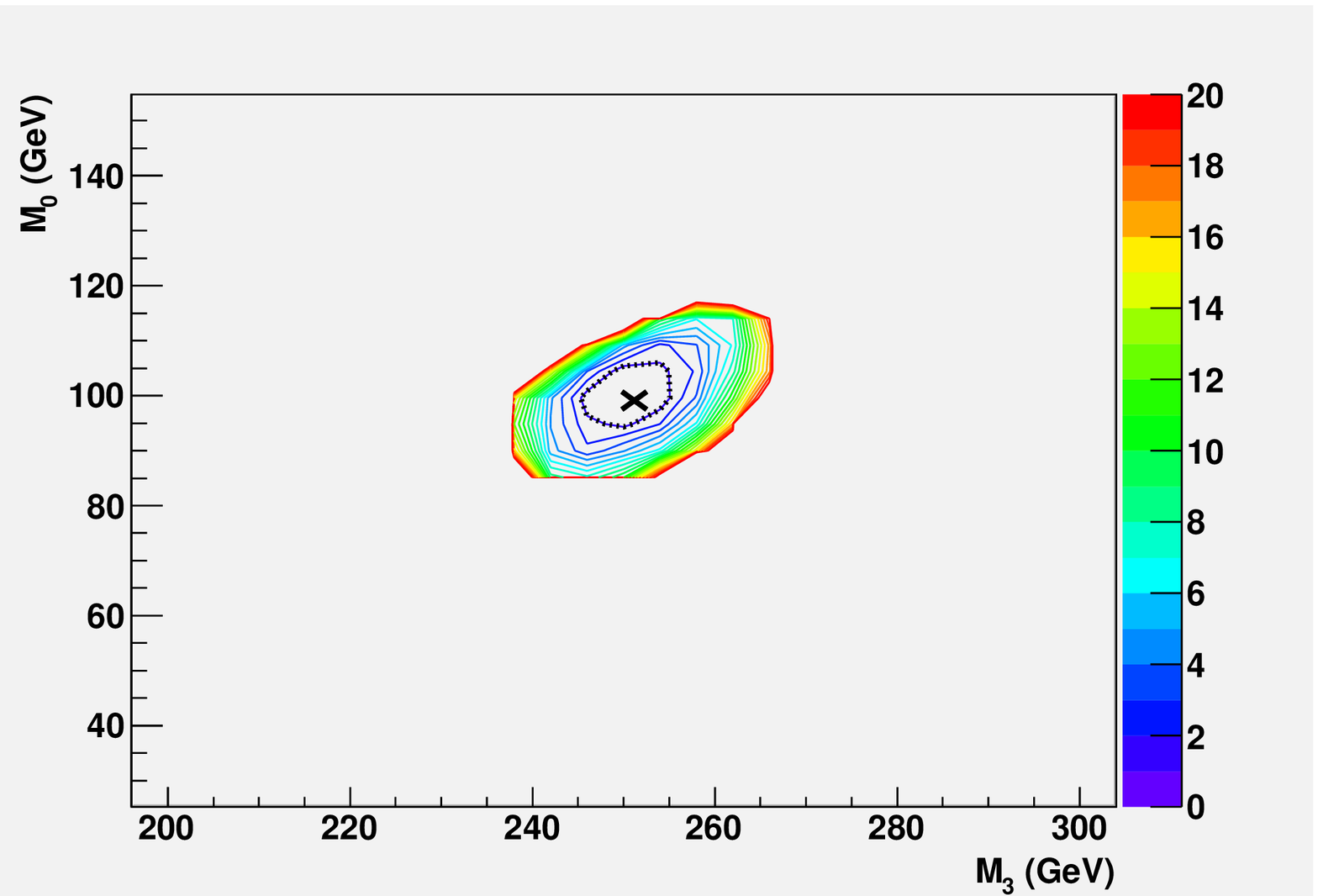}
\\
\end{tabular}
\caption{\label{fig:plot_M0_M3_14_TeV_10_invfb} $\Delta\chi^2$
  contours showing $M_{0}$ against $M_{3}$ for 14~TeV/10~fb${}^{-1}$
  data. Fits are based on the four standard edges of group~I with
  rates (upper right), on the observables of groups~I and II with
  (middle right) and without rates (middle left), and on observables
  of groups~I, II, and III with (lower right) and without rates (lower
  left).}}

Going from 1~fb$^{-1}$ at 7~TeV to 1~fb$^{-1}$ at 14~TeV leads to a
significant reduction of uncertainties. With rates and the observables
of groups~I and II included, we obtain a good fit of the non-universal
model, with errors in the range of 10\% for $M_0$, $M_1$ and $M_2$,
and about 3\% for $M_3$. The fit to $\tan\beta$ has an accuracy of
50\%, compared to 30\% for the case of universal mSUGRA fits. With
10~fb$^{-1}$ at 14~TeV, finally, one can determine $M_3$ with an error
of less then 2\%, while the uncertainties on $M_0$, $M_1$ and $M_2$
are in the range of 5\%. Considering the observables of
groups~I and II together with event rates is sufficient to arrive at
the final LHC accuracy for the GUT-scale gaugino mass parameters
$M_1$, $M_2$ and $M_3$, see Table~\ref{tab:SPS1a_nonuniversal} and
Figure~\ref{fig:plot_M0_M3_14_TeV_10_invfb}.

\clearpage

\section{Conclusions and outlook}
\label{sec:conclusion}

Cross sections and branching ratios provide important information on
the Lagrangian parameters of TeV-scale supersymmetry. We have
presented a new method to include event rates, \ie\ cross
sections$\,\times\,$branching ratios$\,\times\,$cut acceptances, into
global fits of SUSY pa\-ra\-me\-ters at the LHC. While we expect event
rates to be particularly important for SUSY scenarios that are not
well constrained by the measurement of the standard kinematic edges in
cascade decays, we have demonstrated that cross sections and branching
ratios also add important information to fits of SPS1a-type minimal
supergravity models.  In particular, we find that event rates are
crucial for a reliable determination of the mSUGRA parameters in the
initial phase of LHC data taking at 7~TeV collision energy with
1~fb$^{-1}$ integrated luminosity.  We have also studied a more
general class of models where we allow for individual, in general
non-universal, gaugino masses $M_1$, $M_2$ and $M_3$ as fit parameters
instead of a common mass $M_{1/2}$.  The parameter determination of
such models improves significantly when rates are included.

The purpose of this paper is to establish a new method for including
event rates into global SUSY parameter fits at the LHC and to
quantitatively study their impact using SPS1a as a case study. Our
numerical results are based on \Fittino, and the calculation of event
rates will be included in the next release of the \Fittino\ program
package. Note, however, that the method is general and can also be
used with other fitting codes. There are many possible studies that
could follow from the method and the results presented here. We shall,
for example, extend the range of applicability of our cut acceptance
calculation by including three-body decay modes and decays into
on-shell $Z$ and Higgs bosons. Moreover, the estimated uncertainty on
the acceptance should be verified by a detector simulation. So far, we
have focused on LHC data alone, but shall address the interplay of LHC
results with low-energy observables, other collider data and dark
matter constraints in future analyses (\cf~\cite{ Lafaye:2007vs,
  Bechtle:2009ty, Trotta:2008bp, Flacher:2008zq, AbdusSalam:2009qd,
  Buchmueller:2009fn, Akrami:2009hp}). We have seen that event rates
are particularly important to analyze more general models, such as
models with non-universal gaugino masses. We thus plan a more
comprehensive study of the determination of the gaugino mass pattern
from LHC data, which may play a crucial role to distinguish models of
supersymmetry breaking.

\section*{Acknowledgments}
We are particularly grateful to Peter Wienemann for his invaluable
help in all matters related to the technical aspects of {\tt Fittino}.
Thanks also to him, Philip Bechtle and Klaus Desch for many
discussions on physics and fitting. We have appreciated Werner Porod's
help with {\tt SPheno}, and many discussions on experimental aspects
of SUSY searches with Lutz Feld, Niklas Mohr, Daniel Sprenger and
Matthias Edelhoff. This work has been supported in part by the
Helmholtz Alliance ``Physics at the Terascale'', the DFG SFB/TR9
``Computational Particle Physics'', the BMBF ``Verbundprojekt
HEP-Theorie'' under contracts 05H09PDE and 05H09PAE, and the European
Community's Marie-Curie Research Training Network under contract
MRTN-CT-2006-035505 ``Tools and Precision Calculations for Physics
Discoveries at Colliders''. MK would like to thank the CERN TH
division for their hospitality.

\end{document}